\documentstyle[12pt]{article}
\global\parskip 6pt
\begin{document}
\newcommand{\nc}{\newcommand}
\newcommand{\rnc}{\renewcommand}
\catcode`\@=11                                         
\@addtoreset{equation}{section}                        
\rnc{\theequation}{\arabic{section}.\arabic{equation}} 
\nc{\be}{\begin{equation}}
\nc{\ee}{\end{equation}}
\nc{\bea}{\begin{eqnarray}}
\nc{\eea}{\end{eqnarray}}
\rnc{\a}{\alpha}
\nc{\g}{\gamma}
\rnc{\d}{\delta}
\nc{\e}{\eta}
\nc{\f}{\phi}
\nc{\fb}{\bar{\phi}}
\nc{\p}{\psi}
\rnc{\pb}{\bar{\psi}}
\rnc{\c}{\chi}
\nc{\cb}{\bar{\c}}
\nc{\m}{\mu}
\nc{\n}{\nu}
\rnc{\o}{\omega}
\rnc{\t}{\theta}
\nc{\tb}{\bar{\theta}}
\nc{\M}{{\cal M}}                    
\nc{\C}{{\cal A/G}}                  
\nc{\A}[1]{{\cal A}^{#1}/{\cal G}^{#1}}
\nc{\RC}{{\cal R}_{\C}}              
\nc{\RM}{{\cal R}_{\M}}              
\nc{\ra}{\rightarrow}
\nc{\ot}{\otimes}
\rnc{\lg}{{\bf g}}                   
\nc{\cs}{\c_{s}}                     
\nc{\del}{\partial}
\rnc{\O}[1]{\Omega^{#1}(M,\lg)}      
\nc{\h}[1]{{\bf H}_{A}^{#1}}         
\begin{titlepage}
\newlength{\prep}
\settowidth{\prep}{NIKHEF-H/91-???}
\begin{flushright}
\parbox{\prep}{
 \begin{flushleft}
NIKHEF-H/91-28\\
MZ-TH/91-40\\
November 1991
 \end{flushleft}
}
\end{flushright}
\begin{center}
{\Large\bf ${\bf N=2}$ Topological Gauge Theory,}\\
\vskip .1in
{\Large\bf the Euler Characteristic of Moduli Spaces,}\\
\vskip .1in
{\Large\bf and the Casson Invariant}\\
\vskip .3in
{\bf Matthias Blau}\footnote{e-mail: t75@nikhefh.nikhef.nl, 22747::t75} \\
\vskip .10in
{\em NIKHEF-H}\\
{\em P.O. Box 41882, 1009 DB Amsterdam}\\
{\em The Netherlands}
\vskip .15in
{\bf George Thompson}\footnote{e-mail: thompson@vipmzt.physik.uni-mainz.de}\\
\vskip .10in
{\em Institut f\"ur Physik} \\
{\em Johannes-Gutenberg-Universit\"at Mainz}\\
{\em Staudinger Weg 7, D-6500 Mainz, FRG}\\
\end{center}
\begin{abstract}
\noindent We discuss gauge theory with a topological $N=2$ symmetry.
This theory
captures the de Rham complex and Riemannian geometry of some underlying
moduli space $\cal M$ and the partition function  equals the Euler number
$\chi({\cal M})$ of $\cal M$. We explicitly deal with moduli spaces
of instantons and of flat connections in two and three dimensions.
To motivate our constructions we explain the relation between the
Mathai-Quillen formalism and
supersymmetric quantum mechanics and introduce a new kind of supersymmetric
quantum mechanics based on the Gauss-Codazzi equations. We interpret the
gauge theory actions from the Atiyah-Jeffrey point of view and relate them to
supersymmetric quantum mechanics on spaces of connections.
As a consequence of these considerations we propose the Euler number
$\chi({\cal M})$ of the moduli space of flat connections
as a generalization to arbitrary three-manifolds of the Casson
invariant. We also comment on the possibility of constructing a topological
version of the Penner matrix model.
\end{abstract}
\end{titlepage}

\tableofcontents
\setcounter{footnote}{0}
\section{Introduction}

The purpose of this paper is to investigate in some detail the properties
of gauge theories with an $N=2$ topological supersymmetry (models of this type
have appeared previously in \cite{ewtop,bbt,yamron}).
These theories describe
the de Rham complex and Riemannian geometry of some underlying moduli space
$\M$,
in contrast with the
standard $N=1$ gauge theories \cite{ewdon} which model the deformation
complex of $\M$ and capture the geometry
of the Atiyah-Singer \cite{as} universal bundle (see e.g.
\cite{bs,kanno,bbt,pr}).
The most important property of this class of theories is that
formally the partition function of the corresponding
$N=2$ action $S_{\M}$
equals the Euler number of $\M$,
\[ Z(S_{\M})=\c(\M)\;\;,\]
i.e. the Euler characteristic of the de Rham complex of $\M$. That $N=2$
theories may have this property was first suggested by Witten \cite{ewtalk},
who has recently shown \cite{ewwzw} that the twisted Kazama-Suzuki models
\cite{kasu} calculate the Euler number of vector bundles over the moduli
space of Riemann surfaces.

All this is of course quite reminiscent of the properties of
supersymmetric quantum mechanics \cite{ewqm}. That $N=2$ topological gauge
theory is indeed closely related to supersymmetric quantum
mechanics on spaces of connections is seen most clearly within
the framework of the Mathai-Quillen formalism \cite{mq} (as applied to
topological field theories by Atiyah and Jeffrey \cite{aj}).
As pointed out by Atiyah and Jeffrey, this formalism (whose relation
with supersymmetric quantum mechanics we will explain in section 2) can be
used to define some kind of {\em regularized Euler number} $\cs(E)$ of a
vector bundle $E$, depending on a section $s$ of $E$, in cases where
the classical (co)homological or differential geometric definitions are
not terribly useful, e.g. when $E$ is infinite dimensional. Moreover, the
integral expression for $\cs(E)$ can be interpreted as the partition function
of a topological field theory.

In order to illustrate this we will briefly review the classical
Mathai-Quillen formalism (section 2.1) and
then apply it formally (in the spirit of Atiyah and Jeffrey) to the
infinite dimensional loop space $LM$ of a manifold $M$ and its tangent
bundle. We will show that for a special class of sections $s$ the regularized
Euler number $\cs(LM)$ is precisely the partition function of supersymmetric
quantum mechanics and hence, in particular,
equal to the rigorously
defined Euler number $\c(M)$ of $M$ (section 2.2). Conversely, it is possible
to derive the general form of the Mathai-Quillen representative of the Euler
character of a finite-dimensional vector bundle from supersymmetric quantum
mechanics and these two observations allow us to clarify
considerably the meaning of the regularised Euler number of an infinite
dimensional vector bundle.

There is yet another way of obtaining the Euler number of some manifold
from supersymmetric quantum mechanics whose classical counterpart is
based on a combination of the Gauss-Bonnet theorem
with the Gauss-Codazzi equations. These describe the
curvature of some embedded submanifold in terms of the curvature of
the ambient manifold and the extrinsic curvature (second fundamental
form) of the submanifold. The idea is thus to embed the manifold $M$
into some space $Y$ (e.g. Euclidean space $R^{k}$ for $k$ sufficiently
large) whose curvature is known and to combine supersymmetric quantum mechanics
on the latter with a supersymmetric delta function imposing the restriction to
$M$. As this construction appears not to have been discussed in the literature
before, and as it is the prototype of the procedure we will adopt when
considering gauge theories, we explain it in the case of $S^{2}\subset R^{3}$
in section 2.3 (see \cite{btagqm} for the general case).

Having recreated supersymmetric quantum mechanics, which
can be regarded as the simplest
example of a topological field theory \cite{brtnic,bs2,pr}, in this way
it is tempting to apply these ideas to spaces of connections to
construct topological gauge theories. Precisely this has been done by
Atiyah and Jeffrey \cite{aj} who showed that the action of Donaldson theory
\cite{ewdon} can be interpreted as the Mathai-Quillen realization of the
Euler number of an infinite dimensional vector bundle of self-dual two
forms over the orbit space ${\cal A/G}$ of gauge equivalence classes of
connections.

Our main interest here will be in theories where the bundle in question
is (related to) the tangent bundle of ${\cal A/G}$. In this context
Atiyah and Jeffrey have shown that the three-dimensional topological
gauge theories of \cite{ewtop,brt,bg,bbt} can be interpreted as Lagrangian
descriptions of the Euler number $\cs({\cal A/G})$ for the section
$s(A)=*F_{A}$ of $T({\cal A/G})$. In the case that the underlying
three-manifold $M$ is a homology three-sphere (so that the non-trivial
flat connections, the zeros of $s$, are irreducible and isolated) this is
in agreement with Witten's identification of the partition function of
these theories with the Casson invariant \cite{cas} and Taubes' observation
\cite{tau} that the Casson invariant can be interpreted as the Euler number
of ${\cal A/G}$ defined by the vectorfield $*F_{A}$.

This theory is already an $N=2$ model in disguise \cite{ewtop,bbt} and we
will show that it has the feature in common with supersymmetric quantum
mechanics that its partition function can be identified with the Euler number
$\c(\M)$ of some finite dimensional space $\M$, in this case the moduli space
of flat connections. In conjunction with the considerations of \cite{ewtop,tau}
this suggests that also the Casson invariant could
in general be defined as $\c(\M)$. We will come back to this proposal
in section 4.3.

It will become clear in the course of this paper that
there are a number of features peculiar to the case of
flat connections in three dimensions. However,
the construction of $N=2$ actions
with the property that the partition function $Z$ is equal to the
Euler number of some moduli space $\M$ is (formally, i.e.
ignoring analytical questions) completely general and
not limited to this example.
To illustrate this we will also construct these $N=2$ gauge theories
in the somewhat simpler, although perhaps
geometrically less transparent, context of flat connections in two
dimensions and instantons.  The `simpler' here refers to the fact that
the deformation complex is `short' in these examples.

As our proof that $Z =\c(\M)$ will be be
based on the Gauss-Bonnet theorem and the Gauss-Codazzi equations
(i.e. we show
explicitly that $Z$ reduces to an integral over $\M$ of
the exponential of the Riemann curvature ${\cal R}_{\M}$ of $\M$, determined
from the embedding of $\M$ into $\C$) we
will review some of the more elementary aspects of
Riemannian geometry of ${\cal A/G}$ and $\M$ in section 3.1.

We then show (section 3.2)
how to construct Lagrangian descriptions of these geometries
in terms of $N=2$ superfields (see \cite{hor} for the superfield
formulation of topological field theories). The actions will
essentially consist of two parts. One is universal, i.e. common to
all $N=2$ topological gauge theories, and describes the Riemannian
geometry of $\C$. It is the counterpart of the supersymmetric quantum
mechanics action for $Y$ or $R^{k}$ mentioned above.
The other part depends on the choice of moduli space
$\M$. It serves to restrict the theory to $\M\subset\C$ in an $N=2$
invariant way and corresponds to the delta function imposing the restriction
to $M\subset Y$.

As in the $N=1$ theories there is a considerable degree of freedom and
arbitrariness
in the specific choice of Lagrangian. And as the construction of
$N=1$ Witten type topological field theories, Lagrangian realizations
of cohomological field theories defined by intersection theory on
some moduli space, is well understood \cite{lp,bs,brt,ms,bbt}
(and the significance of having a particular Lagrangian realization
at one's disposal should not be overemphasized) we will be rather
brief about these matters here. A fairly detailed analysis of these actions,
in particular with regard to questions of gauge fixing, can be found
in the lectures \cite{gtlec}.

In section 4 we complete the circle of ideas involving $N=2$
topological gauge theories, the Mathai-Quillen formalism and
supersymmetric quantum mechanics. We interpret the topological
actions of section 3.2 from the Atiyah-Jeffrey point of view (section 4.1)
and show that they too can be regarded as Mathai-Quillen realizations
of Euler numbers of certain infinite-dimensional vector bundles,
regularized to give (as in the case of supersymmetric quantum mechanics)
the Euler number of some finite-dimensional space, in this case of the
moduli space $\M$ in question.

In order to understand the emergence and role of de Rham cohomology in
these theories we also explain their relation with
supersymmetric quantum mechanics on $\C$. In particular
we will see that the quantum mechanics theory associated with the space $\C$
of gauge orbits on a three-manifold $M$ (and a particular section of the
tangent bundle of the loop space of $\C$) is nothing other than Donaldson
theory on $M\times S^{1}$, as could have been anticipated from \cite{at}.
In the topologically trivial sector this theory in turn is equivalent to
a three-dimensional topological gauge theory as (in accordance with
general properties of supersymmetric quantum mechanics) only the
time-independent modes contribute to the partition function. As we will
explain in more detail in \cite{btagqm} this theory is, as expected,
precisely the previously constructed $N=2$ gauge theory of flat connections.
This gives an alternative demonstration of the relation between Floer
(instanton) homology and the Casson invariant.

We then turn our attention to the Casson invariant itself (section 4.3).
We review the intersection theory definition of the
Casson invariant \cite{cas} and its relation with Taubes' gauge-theoretic
definition \cite{tau}. We comment on the generalizations
suggested in the mathematics literature and the possibility
of extracting from them a prescription for dealing with reducible connections
in the path-integral. We then make some remarks on the structure of the
moduli space $\M$ of flat connections in three dimensions and, in the light
of this, look at the status of our
suggestion that $\c(\M)$ be regarded as an appropriate generalization of
the Casson invariant to arbitrary three-manifolds.

Finally let us point out that
the property $Z=\c(\M)$ also immediately brings to mind the Penner matrix
model \cite{pen,dv} and suggests the possibility of constructing a
topological version of this theory. This is work in progress \cite{bjnt}
and we will comment on this possibility, as well as on other
possible applications and generalizations, in section 5.

\section{The Mathai-Quillen Formalism and Supersymmetric Quantum Mechanics}

We will now briefly explain the Mathai-Quillen formalism in the
finite dimensional case (see \cite{mq} for details and \cite{aj,pr}
for discussions in the context of topological field theories).
We then explain the relation between the Mathai-Quillen formalism and
supersymmetric quantum mechanics (section 2.2) and describe the
Gauss-Codazzi form of supersymmetric quantum mechanics (section 2.3).

\subsection{The Mathai-Quillen formalism}
We start with some classical material (see e.g. \cite{botu}). Recall
that an oriented $2m$-dimensional real vector bundle $E$ over a manifold
$X$ has an Euler class $e(E)\in H^{2m}(X,{\bf Z})$. If $\dim X = 2m$, this
class can be evaluated  on (the fundamental class $[X]$ of ) $X$ to give the
Euler characteristic (or Euler number)
\[\c(E)=e(E)[X]\;\;.\]
In particular, if $E=TX$, the tangent bundle of $X$, $\c(TX)\equiv\c(X)$ is the
Euler number of $X$. There are two concrete ways of thinking about $\c(E)$.
On the one hand, the Gauss-Bonnet-Chern theorem provides one with an
explicit differential form representative $e_{\nabla}(E)$ of $e(E)$
constructed from the Pfaffian of the
curvature $\Omega=\Omega^{\nabla}$ of a connection $\nabla$ on $E$,
such that
\be
\c(E)=\int_{X} e_{\nabla}(E)\;\;.\label{eq:aj1}
\ee
On the other hand, $\c(E)$ can be computed as the number of zeros of a
generic section $s$ of $E$ (counted with signs),
\be
\c(E)=\sum_{x:s(x)=0}\pm 1 \;\;.\label{eq:aj2}
\ee
If $E=TX$ (and hence $s$ a vector field on $X$)
this is the content of the classical Hopf theorem.
A more general formula,
\be
\c(E)=\int_{X}e_{s,\nabla}(E)\;\;,\label{eq:aj3}
\ee
obtained by Mathai and Quillen, interpolates between the two quite
different descriptions (\ref{eq:aj1}) and (\ref{eq:aj2}). Here
$e_{s,\nabla}$ is a closed $2m$-form on $X$, depending on both a section
$s$ and a connection $\nabla$,
 with the following properties:
if $s$ is the zero section of $E$, then $e_{s,\nabla}=e_{\nabla}$ and
(\ref{eq:aj3}) reduces to (\ref{eq:aj1}); if one replaces $s$ by $ts$,
with $t\in\bf R$, and evaluates (\ref{eq:aj3}) in the limit
$t\rightarrow\infty$ using the stationary phase approximation, (\ref{eq:aj2})
is reproduced. Moreover $e_{s,\nabla}\equiv e_{s}$ (we will suppress the
dependence on the connection $\nabla$ in the following) is the pullback
to $X$ via $s$ of a closed form $U$ on the total space $E$ of the
vector bundle, $e_{s}=s^{*}U$. $U$ is a representative of the Thom class
\cite{botu} of $E$ but, unlike the classical Thom class which has compact
support in the fibre directions, $U$ is Gaussian shaped along the fibres
(cf. (\ref{eq:aj4}) below).

At this point it will be necessary to introduce some more notation: we let
$\xi^{a}$ denote fibre coordinates of $E$,
$\c_{a}$ corresponding
Grassmann odd variables, and $\Omega^{ab}$ the curvature two-form of $E$, and
we regard $E$ as a vector bundle associated to the principal $G$ bundle $P$
with standard fibre $V$, $E=P\times_{G}V$.
Then the Mathai-Quillen form $U$ can be written as a fermionic integral
over the $\c$'s,
\be
U=\pi^{-m}e^{-\xi^{2}}\int d\c\,e^{\c_{a}\Omega^{ab}\c_{b}/4+id\xi^{a}\c_{a}}
\label{eq:aj4}\;\;.
\ee
$U$ is a representative of the Thom class in the $G$-equivariant
cohomology $H_{G}^{2m}(V)$ of $V$ and
can be regarded as a $G$-equivariant form on $P\times V$ whose horizontal part
descends to the Thom form on $E$. In our (somewhat careless) notation
$s^{*}U$ is obtained from $U$ simply by
replacing $\xi$ by $s(x)$. We can now see explicitly that if we take $s$
to be the zero section of the tangent bundle $TX$,
(\ref{eq:aj4}) coincides with the standard Gauss-Bonnet integrand
as the fermionic integral over $\c$ serves to pick out the highest form part
$\sim (\Omega^{ab})^{m}$ of $\exp \c_{a}\Omega^{ab}\c_{b}$ which can then be
integrated over $M$ to yield $\c(E)$. We will explain in section 2.3 how to
obtain the general Mathai-Quillen formula (\ref{eq:aj4}) from supersymmetric
quantum mechanics.

\subsection{Supersymmetric quantum mechanics from the
Mathai-Quillen formalism and vice versa}

In finite dimensions (\ref{eq:aj4}) may perhaps be regarded as an unnecessary
complication since one has the simple classical formula (\ref{eq:aj1}) at
one's disposal. But, as Atiyah and Jeffrey have pointed out, (\ref{eq:aj4})
acquires particular significance in the case of
infinite dimensional bundles where expressions like
(\ref{eq:aj1}) are quite hopeless
but where it may be possible to give
a meaning to (\ref{eq:aj3}) for a suitable choice of section $s$.
(\ref{eq:aj3})
can then be regarded as defining a regularised Euler number $\c_{s}(E)$, which
is however no longer necessarily independent of $s$. If $s$ is a section
canonically associated with $E$, $\c_{s}(E)$ may nevertheless carry interesting
(topological) information. Indeed, in all the examples to be discussed in this
paper we will find that $\cs(E)$ is actually the (rigorously defined) Euler
number of some finite dimensional vector bundle and, as such, certainly
has topological significance
(see e.g. equations (\ref{aj7},\ref{44},\ref{48})). We will see later that
this is a general feature (and, in a way, the quintessence) of the
Mathai-Quillen formalism whenever the zero set of the section $s$ is
finite-dimensional.

Clearly the Mathai-Quillen formalism is closely related to supersymmetric
quantum mechanics. To make this analogy more precise let us consider, as
our first infinite dimensional example, the case where $X$ is the loop space
$X=LM=\{x(t):S^{1}\rightarrow M\}$ of a finite dimensional Riemannian manifold
$M$
and $E$ is its tangent bundle $T(LM)$. The fibre $T_{x}(LM)$ at a loop $x(t)$
can be identified with the space $\Gamma(x^{*}(TM))$ of sections of the
pullback of the tangent bundle of $M$ to $S^{1}$, i.e. with the space of
vector fields on $M$ restricted to the image of the loop $x(t)$. hence
a natural section of the tangent bundle
$T(LM)$ is $s_{0}(x)(t)=\dot{x}(t)$ which we will use to tentatively
define the regularized Euler characteristic $\c_{s}(LM)$ of $LM$.
With this choice of section the exponent in
(\ref{eq:aj4}),
\be
\xi^{2}-\c_{a}\Omega^{ab}\c_{b}/4 - id\xi^{a}\c_{a}\label{eq:aj5}
\ee
(summation over the fibre indices now includes an integration over $t$)
becomes
\be
S(x)=\int_{0}^{\beta} dt [\dot{x}(t)^{2}-\frac{1}{2}
     \pb_{i}(t)R^{ij}_{\;\;kl}(x(t))\p^{k}(t)\p^{l}(t)\pb_{j}(t)
     +2i\pb_{k}(t)\nabla_{t}\p^{k}(t)]\label{aj6}
\ee
Here we have replaced Lorentz by tangent space indices using the vielbein
$e^{a}_{k}$ corresponding to the fibre metric implicit in (\ref{eq:aj5}),
$\pb_{k}=e_{k}^{a}\c_{a}/2$ (this also converts the prefactor in (\ref{eq:aj4})
to $(2\pi)^{-m}$), and we have replaced $dx^{k}(t)$
by the anticommuting variable $\p^{k}(t)$ (so that (\ref{eq:aj3}) will now
also include an explicit integral over $\p^{k}$). But (\ref{aj6}) is nothing
other than the standard action of $N=2$ supersymmetric quantum
mechanics (see e.g. \cite{ewqm,ag})\footnote{There is a slight clash in
notation here. In the literature on supersymmetric quantum mechanics
this model is usually referred to as $N=1$. In the context of topological
field theories, however, it is more convenient and conventional
to count Majorana charges. In the same way the standard $N=1$ topological
gauge theories are the field theoretic cousins of the $N=\frac{1}{2}$
(Dirac operator) model of \cite{ag}.}
The action usually considered is actually slightly more general, depending on
a potential function $W$ on
$M$, and can be obtained from (\ref{eq:aj5}) by choosing, instead
of the above section $s_{0}(x)$, $s_{W}(x)(t)=\dot{x}(t)+W'(x(t))$.
This option will turn out to be essential in our considerations in section
4.2. More generally still one can replace $W'$ by an arbitrary section $V$
of $TM$ (vector field)
and this will allow us to rederive (\ref{eq:aj4}) (valid, after all, for an
arbitrary section $s$ of $E=TM$).

In either case the regularized Euler number of $LM$,
defined via the Mathai-Quillen
formalism, is precisely the partition function of $N=2$ supersymmetric
quantum mechanics on $M$ which, as is well known, is the Euler number
$\c(M)$ of $M$. The standard way of seeing this is to start with the
definition of $\c(M)$ as the Euler characteristic of the de Rham complex
of $M$,
\be
\c(M)=\sum_{k=0}^{2m}(-)^{k}b_{k}(M)\;\;,\label{euler1}
\ee
(here $b_{k}(M)=\dim H^{k}(M,{\bf R})$ is the $k$'th Betti number of $M$)
and to rewrite this as the Witten index
\be
\c(M)=tr(-)^{F}e^{-\beta H}\label{euler2}
\ee
of the Laplace operator $H=\Delta$ on differential forms (or of its
generalization, defined by the twisted exterior derivative
$d_{W}=e^{-W}d e^{W}$ \cite{ewqm}). One then uses the
Feynman-Kac formula to represent this as a supersymmetric path integral
with the action (\ref{aj6}) (or its generalization)
and periodic boundary conditions on the anticommuting variables
$\p^{k}$ (due to the insertion of $(-)^{F}$ in (\ref{euler2})).
Using the $\beta$-independence of (\ref{euler2}) it can be shown that
only the zero modes of the action contribute to the partition function
(the contributions from the non-zero modes cancelling exactly between the
bosonic and fermionic fields), and the evaluation of the remaining finite
dimensional integral then gives the right-hand side of either
(\ref{eq:aj1}) or (\ref{eq:aj2}), i.e. a path integral proof of either
the Gauss-Bonnet
or the Poincar\'e-Hopf theorem. It is interesting to note that these two
rather different classical formulae for $\c(M)$ simply correspond to a
different choice of section (with fixed connection $\nabla$)
in the Mathai-Quillen expression $e_{s,\nabla}(LM)$ for the Euler number
$\cs(LM)$ of the loop space of $M$.

So far we have derived the action of supersymmetric quantum mechanics by
formally applying the Mathai-Quillen formalism to $LM$, and we have
indicated how to rederive
the classical formulae (\ref{eq:aj1},\ref{eq:aj2}) for the Euler number
$\c(M)$ from the resulting action. What is still lacking to
complete the picture is a derivation of the general
(finite dimensional) Mathai-Quillen formula (\ref{eq:aj4})
from supersymmetric quantum
mechanics and this can be done along the following lines. Consider
the quantum mechanics action corresponding to the section
$s(x)(t)=\dot{x}(t)+\alpha V(x(t))$, where $V$ is a vector field on $M$
and $\alpha\in{\bf R}$ is some real parameter. Introduce
a multiplier field $B$ to write the bosonic part of the action
as $\int_{0}^{\beta}(\dot{x}+\alpha V(x))B-B^{2}/2$. The rest of the action is
as in (\ref{aj6}) with the addition of the term
$\pb_{i}\nabla_{k}V^{i}\p^{k}\equiv\pb.dV$
(in our notation we will not distinguish between the vector
field $V$ and its metric dual one-form).
Now
scale $B$ and
$\pb$ by $\beta^{-1/2}$. The contributions from all the non-zero-modes
can again be shown to cancel identically and one is left with an action
of the form
\be
B^{2}+\alpha\beta^{1/2} (VB + \pb.dV) + ({\rm curvature\;terms})\label{SV}
\ee
which - upon rescaling $\alpha$ and integrating over $B$ - reproduces
precisely the Mathai-Quillen formula (\ref{eq:aj3},\ref{eq:aj4}).

In the light of the above let us now make a few more comments concerning
the Mathai-Quillen formalism and the significance of the regularised
Euler number $\cs(E)$ of an infinite dimensional
vector bundle $E$. First of all we want to draw attention to the fact that
what is a section in the Mathai-Quillen formalism is in other contexts
called a Nicolai map. All Witten type
topological field theories have a complete Nicolai map \cite{brtnic,pr}
and it is known that in these theories the partition function can be reduced
to a sum (integral) over the zeros of the Nicolai map. In the present case
this is either ($s_{0}(x)(t)=\dot{x}(t)$) the space $M$ of constant paths
$\dot{x}=0$ or ($s_{W}(x)(t)=\dot{x}(t)+W'(x(t))$) the set of critical points
of $W$. Indeed, by squaring and integrating one sees that $\dot{x}
+W'(x)=0$ implies $\dot{x}=W'(x)=0$ (this we will refer to as the
`squaring argument' in the following). In this case one obtains the Euler
number of $M$ in the form
\be
\c(M)=\sum_{k}\c(M_{W}^{(k)})\;\;,\label{morse}
\ee
where the $M_{W}^{(k)}$ are the connected components of the critical point
set of $W$ and relative orientations are to be taken into account.
In (\ref{morse}) the critical point set of $W$ can also be replaced by
the zero set of any (not necessarily gradient) vector field $V$ on $M$.
The corresponding quantum mechanics action also has a Nicolai map.
The above squaring argument fails, however, as the cross-term $\dot{x}.V$
does not necessarily integrate to zero.
Thus the partition function reduces not to a delta function
onto the critical points but only to a Gaussian
(as in (\ref{eq:aj4},\ref{SV})),
albeit arbitrarily sharply peaked around the critical points of $V$.
The analogue of (\ref{morse}) is then reproduced in the limit
$\alpha\rightarrow\infty$.

Whichever section (action) we use, what we have found is that
the Mathai-Quillen formalism reproduces supersymmetric quantum mechanics
when applied formally to the loop space $LM$ of a finite dimensional
manifold $M$, and that the regularized Euler number of $LM$ (in the sense
of Atiyah and Jeffrey) is
\be
\c_{s}(LM)=\c(M)\label{aj7}\;\;.
\ee
Conversely, we have seen that we can derive the Mathai-Quillen generalization
$e_{s,\nabla}$ (\ref{eq:aj3},\ref{eq:aj4}) of the Gauss-Bonnet integrand
from supersymmetric quantum mechanics.

Equation (\ref{aj7}) shows that (\ref{morse}) is in a way also the essence of
the definition of the regularised Euler number $\c_{s}(LM)$.
One {\em defines} $\c(LM)$ to be equal to the Euler number of the zero set
of some vector field on $LM$. In the finite dimensional case this is,
according to (\ref{morse}), not a definition but an equality. Here, if one
chooses the section of $T(LM)$ to be any of those discussed above
one recovers the result (\ref{aj7}).

This is a general feature of the
Mathai-Quillen formalism in the context of infinite-dimensional bundles:
any `reasonable' definition of the regularised Euler number (any
choice of section with a finite-dimensional zero-set)
will equate it to the Euler number of
some finite-dimensional vector bundle. The latter is, of course, well
defined and unique, while it is the identification of the former with the
latter which is not unique. In fact, there is no good reason for
different choices of sections $s$ making $\cs(E)$ well-defined to
give the same result in general. We will encounter examples of this
in later parts of this paper.

\subsection{The Gauss-Codazzi form of supersymmetric quantum mechanics}

There is yet one more form for the Euler character that can be obtained
from these supersymmetric quantum mechanics models. We could wish to
determine the Euler character of a manifold $M$ by embedding it into an
ambient space $Y$ whose curvature tensor is known (e.g. into Euclidean space
$R^{k}$ with $k$ large enough) and then use the Gauss-Codazzi equations
(cf. below) to determine the curvature tensor of $M$, the Gauss-Bonnet
theorem then giving an explicit result for the Euler character of $M$.

The Gauss-Codazzi equations express the curvature of $M$ in terms of the
curvature of the ambient space $Y$ and the second fundamental form
(extrinsic curvature) of the embedding $i:M\hookrightarrow Y$
of $M$ into $Y$. The second fundamental form of $(M,i)$
is a section $K$ of $Sym^{2}(T^{*}M)\ot N_{M}$ ($N_{M}$
is the normal bundle to $TM$
in $TY|_{M}=i^{*}TY$) defined by
\be
K_{M}(X,Y)=(\nabla_{i_{*}X}i_{*}Y)^{\perp}\;\;,\label{excur}
\ee
where $X,Y\in TM$, $\nabla$ is the Levi-Civit\`a connection on $Y$, and
$(.)^{\perp}:i^{*}TY\ra N_{M}$ the projection. The Gauss-Codazzi
equations\footnote{Actually, the Gauss part of the Gauss-Codazzi equations;
the Codazzi equations express the normal part of the curvature ($W\in N_{M}$
in (\ref{17}) in terms of $K_{M}$ and its derivative.}
now state that in terms of $K_{M}$ and the curvature ${\cal R}_{Y}$ of $Y$
the curvature ${\cal R}_{M}$ of $M$ is given by
\bea
\langle{\cal R}_{M}(X,Y)Z,W\rangle &=& \langle{\cal R}_{Y}(X,Y)Z,W\rangle
\nonumber\\&+&
(\langle K_{M}(Y,Z),K_{M}(X,W)\rangle -(X\leftrightarrow Y))
\label{17}\;\;.
\eea

The supersymmetric quantum mechanics action $S_{M}$
yielding $\c(M)$ in terms of the
integral of (\ref{17}) will itself have a form resembling that of the
Gauss-Codazzi equations. It will consist of the standard action
$S=S_{Y}$ (\ref{aj6}), describing
the curvature of $Y$ and corresponding to the first line of (\ref{17}),
and of a term $S^{0}_{M}$ performing the restriction to $M\subset Y$ in a
supersymmetric way and giving rise to the extrinsic curvature terms.
This idea is easily carried out in general \cite{btagqm}
and will also underly our construction
of gauge theory actions in section 3.2 (with $M\ra\M$ and $Y\ra\C$). Here
we will, for concreteness, consider only the example of $S^{2}$ embedded in
$R^{3}$.

In the case of $Y=R^{3}$ (\ref{aj6}) becomes
\be
S_{Y}(x)=\int dt [2i \dot{x}(t)B(t) + B(t)^{2}
    +i\pb_{k}(t)\partial_{t}\p^{k}(t)]\label{aj6a}\;\; ,
\ee
where we have introduced a multiplier field $B(t)$. The path integral
associated with this action is ill defined being the product of infinity
(due to the presence of $x(t)$ zero modes) and zero (due to the $\p$ and
$\pb$ zero modes). Now we wish to cut out the loop space of $S^{2}$,
$LS^{2}$. To do this we work with $N=2$ superfields,
\be
X^{i}(t,\theta,\bar{\theta})=x^{i}(t)+\theta \p^{i}(t) + \bar{\theta}
\pb^{i}(t) + \theta \bar{\theta} B^{i}(t) \label{aj6b}
\ee
and
\be
b(t,\theta,\bar{\theta}) =  \lambda(t) +  \theta \sigma(t) +
\bar{\theta} \bar{\sigma}(t)  +  \theta \bar{\theta} b(t) \;\;,
\label{aj6c}
\ee
and add to the action (\ref{aj6a}) the following ($M=S^{2}$)
\be
S_{M}^{0}=\int dt d\theta d\bar{\theta} b(t,\theta,\bar{\theta})
(X(t,\theta,\bar{\theta})^{2}\, - \, 1 ) \,. \label{aj6d}
\ee

In terms of components this addition is essentially a delta function
constraint on the paths so that they lie in $LS^{2} \hookrightarrow
LR^{3}$, plus similar constraints on the tangents. By standard arguments
we need only restrict our attention to the zero mode sector of the
theory, and in this limit the partition function becomes
\be
\int d\Phi \, exp[
ib(x^{2}-1) + B^{2} + 2i \sigma x.\pb -2i \bar{\sigma} x.\p +
 2i \lambda (x.B -\p.\pb)] \, , \label{aj6e}
\ee
where $\Phi$ designates all the constant modes. The integral over $b$
restricts us to $S^{2}$ while the integrals over
$\sigma$ and $\bar{\sigma}$ restrict $\p$ and $\pb$ to be tangents to
$S^{2}$. Finally, the $B$ and $\lambda$ integrals yield the exponent
\be
(\p.\pb)( \p.\pb) \, , \label{aj6f}
\ee
which is the curvature term appearing in (\ref{aj6}) with the
Gauss-Codazzi constant curvature that $S^{2}$ inherits from $R^{3}$!

The
upshot of this is that we have indeed recovered the Gauss-Bonnet theorem
{\em with} the explicit Gauss-Codazzi form for the curvature.
In this case the connection with the
Mathai-Quillen construction arises at the level of taking the section
$s_{0} = \dot{x}$ of $T(LR^{3})|_{LS^{2}}$ after the pullback from
$T(LR^{3})$ via $i: S^{2}\hookrightarrow R^{3}$.
This will be explained in more detail in the
gauge theory context in section 4.2.

There is one subtlety in the above prescription which we have ignored
so far. Namely, that the introduction of the constraint (\ref{aj6d}) has
actually lowered the
symmetry of the theory from $N=2$ to $N=1$. The easiest way to see this
is to note that (\ref{aj6a}) is invariant under the simultaneous
interchange $\p\rightarrow \pb$ and $\pb \rightarrow \p$ while
(\ref{aj6d}) has the
symmetry $\p \rightarrow \pb$ and $\pb \rightarrow -\p$ (one also needs
to swap $\sigma$ and $\bar{\sigma}$). Nevertheless the Hamiltonian (which
has no time derivatives) keeps the $N=2$ symmetry of the superfields
manifest. This property will arise again in our analysis of the Euler
character of moduli spaces of flat connections over $3$-manifolds.

\section{${\bf N=2}$ Topological Gauge Theories and the Euler Characteristic
of Moduli Spaces of Connections}

In this section we will construct a topological gauge theory with the property
that its partition function is the Euler characteristic of some
underlying moduli space $\M$. We begin with a brief
review of the Riemannian geometry of the spaces of connections involved
(see e.g. \cite{ims,bv,gp}). We then construct the actions roughly
according to the
recipe explained in section 2.3 without worrying too much about the
geometrical origin of the action and its relation with supersymmetric
quantum mechanics. The connection with the
various ideas of section 2 will be explained in section 4.

\subsection{Riemannian geometry of spaces of connections \ldots}

Let $(M,g)$ be a compact, oriented, Riemannian manifold, $P\ra M$ a principal
$G$ bundle over $M$, $G$ a compact semisimple Lie group and $\lg$ its Lie
algebra. We denote by $\cal A$ the space of (irreducible) connections on $P$,
by $\cal G$ the infinite dimensional gauge group of vertical automorphisms
of $P$ (modulo the center of $G$), by $\O{k}$ the space of $k$-forms on
$M$ with values in the adjoint bundle $ad\,P:=P\times_{ad}\lg$ and by $d_{A}$
the covariant exterior derivative. The spaces $\O{k}$
have natural scalar products defined by the metric $g$ on $M$ (and the
corresponding Hodge operator $*$) and an invariant scalar product $tr$ on
$\lg$, namely
\be
\langle X,Y \rangle = \int_{M}tr(X*Y)\;\;,\;\;\;\;\;\;\;\;X,Y\in\O{k}\;\;.
\label{8}
\ee
The tangent space $T_{A}{\cal A}$ to $\cal A$ at a connection $A$ can be
identified with $\O{1}$. At each point $A\in\cal A$, $T_{A}{\cal A}$ can be
split into a vertical part $V_{A}=Im(d_{A})$ (tanget to the orbit of
$\cal G$ through $A$) and a horizontal part $H_{A}=Ker(d_{A}^{*})$ (the
orthogonal complement of $V_{A}$ with respect to the scalar product (\ref{8})).
Explicitly this decomposition of $X\in\O{1}$ into its vertical and horizontal
parts is
\be
X=d_{A}G_{A}^{0}d_{A}^{*}X + (X-d_{A}G_{A}^{0}d_{A}^{*}X)\;\;,\label{9}
\ee
where $G_{A}^{0} = (d_{A}^{*}d_{A})^{-1}$ is the Greens' function of the
scalar Laplacian (which exists if $A$ is irreducible).

Working with
appropriate Sobolev spaces of connections (we will not indicate this explicitly
in the following) it can be shown that the space $\C$ of gauge equivalence
classes $[A]$ of connections is a smooth Hausdorff Hilbert manifold.
It is often convenient to
identify the tangent space $T_{[A]}\C$ with $H_{A}$ for some representative
$A$ of the gauge equivalence class $[A]$. $\cal G$
acts on $\cal A$ isometrically and preserving the above decomposition so that
(\ref{8}) induces a metric on $\C$ making the principal projection
$\pi:{\cal A}\ra\C$ a Riemannian submersion.

The Riemannian curvature of $\C$
can now be computed straightforwardly in a variety
of ways and is
\bea
\langle\RC(X,Y)Z,W\rangle&=&
\langle *[\bar{X},*\bar{W}],G_{A}^{0}*[\bar{Y},*\bar{Z}]\rangle
-\langle *[\bar{Y},*\bar{W}],G_{A}^{0}*[\bar{X},*\bar{Z}]\rangle\nonumber\\
& &+2 \langle *[\bar{W},*\bar{Z}],G_{A}^{0}*[\bar{X},*\bar{Y}]\rangle
\label{10}
\eea
(with $W,X,Y,Z\in T_{[A]}\C$ and $\bar{W},\bar{X},\bar{Y},\bar{Z}$ local
horizontal extensions of their lifts to $H_{A}$).

We now turn our attention to certain finite dimensional (moduli) subspaces
$\M$ of $\C$. Obvious examples that come to mind are moduli spaces of
flat connections ($F_{A}\equiv dA+\frac{1}{2}[A,A]=0$), Yang-Mills connections
($d_{A}^{*}F_{A}=0$) and (in dimension 4) instantons ($P_{+}F_{A}\equiv
\frac{1}{2}(1+*)F_{A}=0$). Following Groisser and Parker \cite{gp}, who
discussed the case of instantons, we will describe the Riemannian curvature
$\RM$ of $\M$ in terms of $\RC$ (\ref{10}) and the second fundamental form
(extrinsic curvature) of the embedding $i:\M\hookrightarrow\C$, using the
Gauss-Codazzi equations (cf. section 2.3).
At this point it is rather awkward to continue
in this generality and we will therefore deal explicitly now with the moduli
spaces $\M_{2}$ and $\M_{I}$ of flat connections in two dimensions and
instantons. Afterwards we will treat the moduli space $\M_{3}$.

The formal (Zariski) tangent space $T_{[A]}\M_{2}$ ($T_{[A]}\M_{I}$)
can be identified with the subspace of
$T_{[A]}\C\sim H_{A}=\{X\in\O{1}:d_{A}*X=0\}$
satisfying the linearized equations $d_{A}X=0$ (respectively $P_{+}d_{A}X=0$).
Put differently, one has $T_{[A]}\M\sim\h{1}$ where $\h{k}$ is the $k$-th
cohomology group of the flat connection or instanton deformation complex,
\be
0\ra\O{0}\stackrel{d_{A}}{\ra}\O{1}\stackrel{d_{A}}{\ra}\O{2}\ra 0 \label{11}
\;\;,
\ee
\be
0\ra\O{0}\stackrel{d_{A}}{\ra}\O{1}\stackrel{P_{+}d_{A}}{\ra}\O{2}_{+}\ra 0
\label{12}
\ee
(note that $d_{A}^{2}=0$ ($P_{+}d_{A}^{2}=0$) for $A$ flat (an instanton)).
Although we shall not be too concerned with the analytical properties of these
moduli spaces (see, however, section 4.3 for remarks on the structure of the
not so well understood moduli spaces $\M_{3}$)
we mention that the zeroth cohomology groups of (\ref{11})
and (\ref{12}) as well as the second cohomology group of (\ref{12}) (by
Poincar\'e duality) are zero for $A$ irreducible, and that the second
cohomology group $\h{2}={\rm Ker}P_{+}d_{A}(P_{+}d_{A})^{*}$ of (\ref{12})
can be shown to be zero at irreducible connections
for particular \cite{ahs} and generic \cite{fu}
metrics. This allows to establish the local smoothness of $\M$ in the
neighbourhood of irreducible connections via the implicit function theorem.
In particular, at smooth points of $\M$ the dimension of $\M$, the dimension
of $\h{1}$ and the index of (\ref{11},\ref{12}) all agree. For more information
see \cite{dk,gol,pr}.

Vectors in $T_{[A]}\M$ can be represented by elements $\bar{X}_{A}\in\O{1}$
of the form
\bea
\bar{X}_{A} &=& X_{A}-d_{A}^{*}G_{A}^{2}d_{A}X_{A}\label{13}\;\;,\\
\bar{X}_{A} &=& X_{A}-(P_{+}d_{A})^{*}G_{A}^{2}P_{+}d_{A}X_{A}\label{14}\;\;,
\eea
where $X_{A}\in H_{A}$ and $G_{A}^{2}$ is the Green's function of the
second Laplacian $d_{A}d_{A}^{*}$ ($(P_{+}d_{A}(P_{+}d_{A})^{*}$) of the
deformation complex (\ref{11},\ref{12}). Indeed, one easily verifies that e.g.
$\bar{X}_{A}$ as given by (\ref{13}) satisfies
$d_{A}\bar{X}_{A}=d_{A}^{*}\bar{X}_{A}=0$. Using (\ref{13},\ref{14}),
the extrinsic curvature $K_{\M}$ can be computed to be
\bea
K_{\M}(X,Y)&=&-d_{A}^{*}G_{A}^{2}[\bar{X},\bar{Y}]\label{15}\;\;,\\
K_{\M}(X,Y)&=&-(P_{+}d_{A})^{*}G_{A}^{2}P_{+}[\bar{X},\bar{Y}]\label{16}\;\;.
\eea
Together with the Gauss-Codazzi equation (\ref{17}),
equation (\ref{10}) and
\bea
\langle K_{\M}(Y,Z),K_{\M}(X,W)\rangle&=&
\langle [\bar{Y},\bar{Z]},G_{A}^{2}[\bar{X},\bar{W}]\rangle\label{18}
\;\;,\\
\langle K_{\M}(Y,Z),K_{\M}(X,W)\rangle&=&
\langle P_{+}[\bar{Y},\bar{Z]},G_{A}^{2}P_{+}[\bar{X},\bar{W}]\rangle\label{19}
\;\;,
\eea
this allows us to express the Riemann curvature tensor of $\M_{2}$ and
$\M_{I}$ entirely in terms of the Green's functions
$G_{A}^{0}$ and $G_{A}^{2}$. And it is in precisely this form that we will
derive $\RM$ from a suitable Lagrangian in the next section.

So far we have dealt with two examples of moduli spaces of connections
whose deformation complices (\ref{11},\ref{12}) are `short' and where
the obstructing cohomology groups $\h{k},k\neq 1$ are (generically) zero.
This will of course not always be the case and one may wonder how much
of the above nevertheless remains valid under more general circumstances.
For concreteness let us consider the moduli space $\M_{3}=\M_{3}(M,G)$
of flat $G$-connections on a three-manifold $M$. Its deformation
complex is (like (\ref{11})) the twisted de Rham complex of $M$,
\be
0\ra\O{0}\stackrel{d_{A}}{\ra}\O{1}\stackrel{d_{A}}{\ra}\O{2}
         \stackrel{d_{A}}{\ra}\O{3}\ra 0 \label{20}\;\;.
\ee
In this case, however, significantly less information can be extracted from it
than in the examples discussed previously. In particular, although the
formal tangent space $T_{[A]}\M$ can still be identified with $\h{1}$, the
index of (\ref{20}) (being zero by Poincar\'e duality) provides no information
on the dimension of $\M$. The cohomology groups $\h{0}$ and $\h{3}$ of
(\ref{20}) are zero at irreducible connections, but there will certainly be
no vanishing theorem for $\h{2}$ in general as $\h{2}\sim\h{1}$ and
$\h{1}\neq 0$ is a necessary condition for having a non-zero-dimensional
moduli space. General results on the structure of the smooth and
singular parts of $\M_{3}$ appear not to be known.
In section 4.3 we shall mention some partial information that can be
extracted from the existing literature
on representation varieties of finitely generated groups.
In the following we
simply assume that we are working with the smooth part of $\M_{3}$ only.

Turning now to Riemannian geometry
let us first look at the analogue of (\ref{13}) in the present case.
Although the Laplacian on two-forms,
$\Delta_{A}^{2}=d_{A}^{*}d_{A}+d_{A}d_{A}^{*}$,
whose Green's function enters into (\ref{13}), has a non-trivial kernel,
(\ref{13}) makes sense in three dimensions as it stands since $\Delta_{2}$
is certainly invertible on ${\rm Im}d_{A}$ (hence in particular on
$d_{A}X_{A}$). Thus equations (\ref{15}) and (\ref{18}) also remain
valid provided that one thinks of $G_{A}^{2}$ as including a projection
onto the orthogonal complement (with respect to (\ref{8})) of $\h{2}$
in $\O{2}={\rm Im}d_{A}\oplus{\rm Im}d_{A}^{*}\oplus\h{2}$. In conclusion
we see that despite the additional complications present in the case
of flat connections in three (and higher) dimensions,
(\ref{15}) and (\ref{18}) remain valid and the expression for the
Riemann tensor $\RM$ is formally identical to that given above for the
moduli space $\M_{2}$.

\subsection{\ldots and its Lagrangian realization}

We will now explain how to construct the topological action $S_\M$ capturing
the geometry of some moduli space $\M$ described in the previous section.
In analogy with the Gauss-Codazzi equation (\ref{17}) and the considerations
of section 2.3 $S_\M$ will
essentially consist of two parts. One of them, $S_\C$, is universal, i.e.
common to all gauge theories with a topological $N=2$ symmetry, and describes
the Riemannian geometry of $\C$ (much in the same way as $N=1$ topological
gauge theories all have a part in common which describes the geometry of the
Atiyah-Singer \cite{as} universal bundle \cite{bs,kanno,bbt,pr}).
The other, $S_{\M}^{0}$ (corresponding to the extrinsic curvature contribution
in (\ref{17})) will depend on the particular moduli space chosen.

We introduce an $N=2$ superconnection \cite{ewtop,bbt}
\be
\hat{A}=A_{\mu}(x,\t,\tb)dx^{\mu}+A_{\t}(x,\t,\tb)d\t +
           A_{\tb}(x,\t,\tb)d\tb\label{21}
\ee
with components
\bea
A_{\mu}(x,\t,\tb) &=& A_{\mu}(x)+\t\psi_{\mu}(x)+\tb\pb_{\mu}(x)+
      \t\tb \Sigma_{\mu}(x)\nonumber\\
A_{\t}(x,\t,\tb) &=& \xi(x)+\t\phi(x)+\tb\rho(x)+\t\tb\eta(x)\nonumber\\
A_{\tb}(x,\t,\tb) &=& \bar{\xi}(x)+\t\bar{\rho}(x)+\tb\fb(x)+\t\tb\bar{\eta}(x)
\label{22}
\eea
In this formulation the fields carry a natural trigrading (a,b,c),
where the first entry is the conventional form degree while the second
and third entries
correspond to the $\theta$ and $\bar{\theta}$ weights respectively.
For our purposes, however, it will be sufficient and more convenient
to assign a bigrading to the fields in such a way that $A=A_{\mu}(x)dx^{\mu}$,
$\t$ and $\tb$ are $(1,0)$-,$(0,-1)$-, and $(0,1)$-forms respectively.
This determines e.g. $\p = \p_{\mu}dx^{\mu}$ and $\f$ to be $(1,1)$- and
$(0,2)$-forms, which is
in agreement with the ghost number assignments
of the standard $N=1$ theories.

{}From $\hat{A}$ we can construct the supercurvature form $\hat{F}$ as
\be
\hat{F}= \hat{d}\hat{A}+\frac{1}{2}[\hat{A},\hat{A}]\label{23}
\ee
($\hat{d} = dx^{\mu}\del_{\mu}+d\t\del_{\t}+d\tb\del_{\tb}$), which transforms
homogeneously ($\d\hat{F}=[\hat{F},\hat{\Lambda}]$) under the supergauge
transformation
\be
\d\hat{A}=\hat{d}\hat{\Lambda}+[\hat{A},\hat{\Lambda}] \;\;.\label{24}
\ee
We will use this supergauge transformation to set
$\xi=\bar{\xi}=\rho-\bar{\rho}
=0$. This reduces (\ref{24}) to the ordinary gauge symmetry
($\hat{d}\hat{\Lambda}= d\hat{\Lambda}$) which we will keep manifest
together with the $N=2$ symmetry. As a consequence of the above
gauge choice the $N=2$ generators $s$ and $\bar{s}$ are now
the superspace derivatives $\del_{\t}$ and $\del_{\tb}$ supplemented
by field dependent gauge transformations. For instance, instead of
$sA=\p,s\p=0$ one now has $sA=\p,s\p=-d_{A}\f$, which is the familiar
equivariant supersymmetry of Donaldson theory \cite{ewdon,kanno}
and any other $N=1$ topological gauge theory.

We pause here to explain
the relationship between the equivariant and non-equivariant versions of
topological field theories (for any $N$). By standard gauge covariance
arguments the calculation of any gauge invariant observable does not
depend on the gauge chosen. If one therefore calculates expectation
values of observables that are both shift supersymmetric and gauge
invariant, in either version of the theory the results will necessarily
be the same. In particular, for the theory at hand the partition
function may be
viewed as the expectation value of $1$, which is certainly gauge and
shift invariant.
The alternative of keeping
the complete set of fields (plus the $N=2$ multiplets
of Faddeev-Popov ghosts required to gauge fix the symmetry (\ref{24})) has
been worked out in \cite{gtlec} and the results of course agree with
those presented here.

{}From the supercurvature $\hat{F}$ with components
\bea
F_{0}&\equiv&\frac{1}{2}F_{\m\n}(x,\t,\tb)dx^{\m}dx^{\n}\nonumber\\
   &=&F_{A}-\t d_{A}\p -\tb d_{A}\pb +\t\tb(d_{A}\Sigma + [\p,\pb])
   \label{25}\\
F_{\t}&\equiv& F_{\m\t}(x,\t,\tb)dx^{\m}\nonumber\\
   &=&(\p -\t d_{A}\f + \tb(\Sigma-d_{A}\rho)+\t\tb(d_{A}\bar{\e}+[\f,\pb]
      -[\rho,\p]))d\t\label{26}\\
.. & & \ldots\nonumber
\eea
we can construct topological actions from the various
contributions to the super Yang-Mills action
$\int_{(M,\t,\tb)}\hat{F}\star\hat{F}$ (where $\star$ is a suitably defined
super Hodge operator). As in conventional $N=1$ topological theories
most of these terms are inessential and have no influence on the
dynamics or calculation of correlation functions. The one term that will
be important for us is
\be
S_{\C}=\int_{M}d\t d\tb F_{\t}*F_{\tb}\label{27}
\ee
which provides propagators for all the components of
$A_{\t}$ and $A_{\tb}$.
The `dynamics' for $A_{\m}(x,\t,\tb)$, which
of course depends on the choice of $\M$, will be specified in terms of
$F_{0}$ or $A(x,\t,\tb)$ (cf. (\ref{32}) and (\ref{38}) below).
$S_{\C}$ is manifestly gauge invariant and $N=2$ supersymmetric and given
explicitly in terms of components by
\bea
S_{\C}&=&\int_{M}d_{A}\f *d_{A}\fb-d_{A}\rho *d_{A}\rho + \eta d_{A}*\p
         +\bar{\eta}d_{A}*\pb\nonumber\\
      &+&\f[\pb,*\pb]+\fb[\p,*\p]-2\rho[\p,*\pb]+\Sigma*\Sigma\label{28}\;\;.
\eea
Let us now analyze this action. The equations of motion of $\eta$ and
$\bar{\eta}$ tell us that
$d_{A}*\p=d_{A}*\pb=0$ so that $\p$ and $\pb$ can be interpreted as horizontal
tangent vectors to $\cal A$, i.e. elements of $H_{A}$ (cf. section 3.1).
The Gaussian integral over $\rho$ generates a term
$[\p,*\pb]G_{A}^{0}*[\p,*\pb]$, and similarly the integrals over
$\f$ and $\fb$ contribute the term $[\pb,*\pb]G_{A}^{0}*[\p,*\p]$.
Away from reducible connections there will be no
scalar zero modes to worry about
so that effectively the action $S_{\C}$ now takes the form
\be
S_{\C}=\int_{M}([\p,*\pb]G_{A}^{0}*[\p,*\pb]+[\pb,*\pb]G_{A}^{0}*[\p,*\p]
     +\Sigma*\Sigma)\label{29}
\ee
where $\p$ and $\pb$ are gauge fixed pointwise at $A$, i.e. satisfy
$d_{A}*\p=d_{A}*\pb=0$.
The important observation is now that the first
two terms of (\ref{29}) are precisely the combinations of Green's functions
appearing in the expression (\ref{10}) for the Riemann tensor $\RC$ so that
(by slight abuse of notation) we can rewrite (\ref{29}) more succinctly as
\be
S_{\C}=\RC+\int_{M}\Sigma*\Sigma\;\;.\label{30}
\ee
It is in this sense that the universal contribution $S_{\C}$ to the action
of any $N=2$ topological gauge theory captures the Riemannian geometry of
$\C$.

We will now explain how to construct the $\M$-dependent part $S_{\M}^{0}$
of the action.
The role of $S_{\M}^{0}$ is to restrict the gauge fields to $\M\subset\C$.
Performing this restriction in an $N=2$ invariant
way automatically provides the extrinsic curvature contribution to the
Gauss-Codazzi equation (\ref{17}).

We will first discuss the example $\M_{2}$ of flat connections in two
dimensions. The obvious way to impose the condition $F_{A}=0$ in the
path integral is to introduce a term $\sim\int_{M}BF_{A}$ into the action
where $B$ is a scalar field. In order to do this in a manifestly $N=2$
invariant way we introduce a scalar superfield
\be
\hat{B} = u + \t\c +\tb\cb +\t\tb B \label{31}
\ee
and consider the action
\be
S_{\M}^{0}=\int_{M} d\t d\tb \hat{B}F_{0}\label{32}
\ee
($F_{0}$ is the two-form part of the supercurvature as given in (\ref{25})).
Written out in components (\ref{32}) is
\be
S_{\M}^{0}=\int_{M} BF_{A}-\c d_{A}\pb +\cb d_{A}\p +u(d_{A}\Sigma +[\p,\pb])
\label{33}\;\;.
\ee
We see that the
$B$-integral (or equation of motion) forces $A$ to be flat while the
$\c$- and $\cb$-integrals tell us that $\p$ and $\pb$ satisfy the
linearized flatness conditions $d_{A}\p=d_{A}\pb =0$. Together with
the previously established conditions $d_{A}*\p=d_{A}*\pb=0$ this means
that the solutions $\p_{A}$ and $\pb_{A}$ to these equations represent
elements of $T_{[A]}\M_{2}$.

Adding $S_{\M}^{0}$ to $S_{\C}$ with an arbitrary coefficient $\a$,
\be
S_{\M}=S_{\C}+\a S_{\M}^{0}\label{34}\;\;,
\ee
(this action should of course still be supplemented by gauge fixing terms
which we will, however, not write explicitly)
and integrating over $\Sigma$ and $u$ one finds that the only contribution
from $S_{\M}^{0}$ to the action $S_{\M}$ (in addition to the above constraints
on the fields $A$, $\p$ and $\pb$) is the $\a$-{\em independent}
term $-[\p,\pb]*G_{A}^{2}[\p,\pb]$ (it can be checked that $\a$ also drops out
of the measure, as it should by supersymmetry).
This is precisely the term (\ref{18})
quadratic in the extrinsic curvature which appears in the Gauss-Codazzi
equation (\ref{17}). Thus what we have achieved so far in this section can
(with the same {\it caveat}
as that preceding (\ref{30})) be summarized by the
equation
\be
S_{\M}(A,\p,\pb)=\RM \label{35}
\ee
with $A$ representing a point $[A]$ of $\M_{2}$ and $\p=\p_{A}$ and
$\pb=\pb_{A}$ representing tangents to that point.

The evaluation of the partition function
\be
Z(S_{\M})=\int_{[A]\in\M}D[A]\int D\p_{A} \int D\pb_{A}
         e^{iS_{\M}(A,\p_{A},\pb_{A})}\label{36}
\ee
is now straightforward and proceeds exactly as in the case of
supersymmetric quantum mechanics \cite{ag,fw}. There are an equal number
$d(\M)=\dim\M_{2}$ of $\p$ and $\pb$ Grassmann-odd zero modes which can be
soaked up, provided that $d(\M)$ is even, by expanding the exponential
to $(d(\M)/2)$-th order. The remaining integral is then of the Gauss-Bonnet
form $\int_{\M} \RM^{d(\M)/2}$. If, on the other hand, $d(\M)$ is odd
(this does not occur for two-dimensional surfaces)
the partition function is zero. Moreover,
if $\M$ is not connected (as will frequently be the case) then care has to
be taken with the relative signs of the contributions from the connected
components of $\M$.

In any case one finds (possibly up
to a numerical factor depending only on the dimension but not on the nature
of $\M$)\footnote{Following the analysis of the normalization of the zero
mode integrals in \cite{ag,fw} this factor can be shown to be $1$.} that
the partition function of the $N=2$ topological gauge theory defined by the
action (\ref{34})=(\ref{28})+(\ref{33}) is the Euler characteristic of
the moduli space $\M_{2}$,
\be
Z(S_{\M})=\c(\M_{2})\label{37}\;\;!
\ee

In the case of instantons all that needs to be changed in the above derivation
is to replace the scalar superfield $\hat{B}$ by a selfdual two-form
superfield $\hat{B}_{+}$ (giving rise to (\ref{19}) instead of (\ref{18})).
And for flat connections in $n>2$ dimensions one could use the action
(\ref{32}) with $\hat{B}$ an $(n-2)$-form. In that case, however, additional
gauge fixing terms are required because of the
`Bianchi' symmetry $\d B_{n-2}=d_{A}\Lambda_{n-3}$ of the action
$\int_{M} B_{n-2}F_{A}+\ldots$. This is a manifestation of the extended length
of the deformation complex (the twisted de Rham complex in $n$ dimensions)
and will give rise to a plethora of new zero modes (corresponding to the
cohomology groups $\h{k},k>1$ of the deformation complex). Nevertheless,
ignoring these complications one will then formally find
$Z(S_{\M})=\c(\M_{n})$ as well. But it should be borne in mind that this
result is on a much less secure footing than (\ref{37}) where the zero modes
are under control and their significance (reducibility) is well understood.

In three dimensions, however, another procedure is available, technically
because of the fact that the required multiplier $B$ is a one-form so that
it can be incorporated into the $N=2$ multiplet of $A$. This is a feature
shared by topological gauge theories based on the moduli space of Yang-Mills
connections {\em in any dimension} \cite{bbt} because
(like $F_{A}$ in $n=3$ dimensions)
$d_{A}*F_{A}$  is an $(n-1)$-form. This also means that the `obvious'
$N=1$ topological theories associated with these moduli spaces will
automatically have an underlying $N=2$ symmetry.

Thus in three dimensions, instead of introducing $\hat{B}$ and using the
analogue of (\ref{32}), one can use a super Chern-Simons action \cite{ewtop}
and we will choose
\bea
S_{\M}^{0}&=&\frac{1}{2}\int_{M}d\t d\tb A(x,\t,\tb)dA(x,\t,\tb)+
             \frac{2}{3}A(x,\t,\tb)^{3} \nonumber\\
          &=&\int_{M} BF_{A} + \pb d_{A}\p\label{38}
\eea
(we have now called the $\t\tb$-component of $A(x,\t,\tb)$ $B$ instead
of $\Sigma$ as it plays the role the multiplier field $B_{n-2}$ plays
in the formulation (\ref{32})). For the theories based on the Yang-Mills
moduli space the appropriate action would of course have been the
super Yang-Mills action $\int_{M}d\t d\tb F_{0}*F_{0}$ (cf. (\ref{25})).

The classical equations of motion that one obtains from the action
(\ref{38}) are
(none too surprisingly) $F_{0}=0$. In particular
the gauge fields are flat while the classical $\p$ and
$\pb$ configurations represent, as above, tangents to the moduli
space $\M_{3}$.
Note that in addition to the $A$, $\p$ and $\pb$ zero modes we will have
an equal number of $B$ zero modes as the solution to the equation of
motion $d_{A}B+[\p,\pb]=0$ will only be unique up to the addition of
one forms $X\in\O{1}$ satisfying the linearized flatness equation
$d_{A}X=0$.

The above considerations have to be modified slightly once we add
(\ref{38}) to the action $S_{\C}$ (\ref{30}). The only essential
modification is that the $\Sigma^{2}\equiv B^{2}$
term in the action $S_{\C}$ now implies that the path integral
is only Gaussian (and not delta-function) peaked around the moduli
space $\M_{3}$ (although this Gaussian can be arbitrarily close to a
delta-function since the coefficient $\a$ of $S_{\M}^{0}$ is arbitrary).
This will be reflected in the fact that the quantum
fluctuations of $A$, which did not make any appearance in the
delta-function examples discussed previously, will play an important role
in the analysis below. The presence of the $B^{2}$ term also ensures that
the $B$ zero modes are damped in the path integral, only contributing to
its overall normalization - a rather welcome feature at this point as
all the relevant
geometry is already encoded into the zero modes of $A$, $\p$ and
$\pb$ and undamped Grassmann even
$B$ zero modes would have just been a nuisance.

With these remarks in mind let us now complete the analysis of this action.
We expand $A$, $\psi$ and $\pb$ about their classical configurations
\be
A=A_{c}+A_{q}\;\;,\;\;\psi=\psi_{c}+\psi_{q}\;\;,\;\;
\pb=\pb_{c}+\pb_{q}\;\;.\label{39}
\ee
The only terms of interest in the action are
\be
\RC +\int_{M}(B*B+\a Bd_{A_{c}}A_{q}+\a [\pb_{c},\p_{c}]A_{q})\;\;.
\label{40}
\ee
Integrating over the $B$ and $A_{q}$ fields
allows us to write the action in its final form
\be
S_{\M}=\RC+\int_{M}[\pb,\p]* G^{2}_{A_{c}}[\pb,\p] = \RM\;\;,\label{41}
\ee
leading, as above, to the result
\be
Z(S_{\M})=\c(\M_{3})\;\;.\label{42}
\ee

\section{Geometry of ${\bf N=2}$ Topological Gauge Theories}

In the previous section we discussed the geometry of moduli spaces of
connections and and constructed topological actions which describe this
geometry. At this point, however, it may not yet be clear why these actions
do the job.
By linking the construction of section 3.2 with the ideas of section 2, we will
now try to provide a more geometric explanation of the origin of these actions.
In particular, we will interpret these actions from the Atiyah-Jeffrey
point of view as being Mathai-Quillen representatives of regularized Euler
numbers of certain infinite-dimensional bundles. These we can then,
according to the calculations of section 3.2, identify with some
finite-dimensional Euler number (as in the case of supersymmetric
quantum mechanics, cf. (\ref{aj7})). Although this interpretation `explains'
the actions to a certain extent, it does not immediately
shed any light on the question in which way these theories can
be regarded as arising from some
infinite dimensional supersymmetric quantum mechanics theories,
and this we will try to remedy in section 4.2 and, from a more computational
point of view, in \cite{btagqm}.
These considerations will naturally
lead us to the generalization of the Casson invariant mentioned in the
introduction (section 4.3).

\subsection{The Atiyah-Jeffrey interpretation}

Here the idea is to show that the action $S_{\M}=S_{\C}+S_{\M}^{0}$
(\ref{28},\ref{33},\ref{38})
has the form of the Mathai-Quillen exponent (\ref{eq:aj5})
for a suitable bundle $E$ and section $s$. This can be done in either of
two ways: by exploiting the geometry  of the principal fibration
$P\ra X$ (${\cal A}\ra\C$) and its associated vector bundles to manipulate
(\ref{eq:aj5}) into the form (\ref{28})+(\ref{33},\ref{38})
with the complete field
content; or (more simply, but also less elegantly) by reducing $S_{\M}$ to
the form (\ref{eq:aj5}). The former has been explained in great detail by
Atiyah and Jeffrey in the case of Donaldson theory and the three-dimensional
theory of flat connections discussed above. For simplicity we will focus on the
latter here, which essentially amounts to performing the manipulations of
section 3.2 (or, equivalently, those leading to (\ref{SV}) in the case of
quantum mechanics).

We begin with the three-dimensional theory. This is the richest of the
models discussed in section 3 and also geometrically the most transparent
(reflected in the fact that no auxiliary fields were required in the
construction of the action). Recalling (\ref{30}) and (\ref{38}) we see
that we can already write $S_{\M}$ in reduced form as
\bea
S_{\M}&=&\RC+\int_{M}BF_{A}+\pb d_{A}\p+B*B\nonumber\\
      &=&-\frac{1}{4}\int_{M}F_{A}*F_{A} + \RC +\int_{M}\pb d_{A}\p\;\;.
\label{43}
\eea
This is precisely of the form (\ref{eq:aj5}), i.e. of the form
\[-\xi^{2}+\c_{a}\Omega^{ab}\c_{b}/4 + id\xi^{a}\c_{a}\;\;,\]
for $s(A)=*F_{A}\in\O{1}$ provided that we rescale $\pb$ appropriately
(to identify the third term of (\ref{43}) note that $\d F_{A}=d_{A}\d A$
and that $\pb$ plays the role of $\c$). $s$ is a section of the tangent bundle
$T{\cal A}$ of $\cal A$ which passes down to a section of $T(\C)$ as
$d_{A}^{*}*F_{A}=0$ by the Bianchi identity $d_{A}F_{A}=0$. In fact,
$s(A)=*F_{A}$ is the gradient vectorfield of the Chern-Simons functional
and as such enters into the
definition of Floer cohomology \cite{floer} as well as into Taubes'
interpretation of the Casson invariant \cite{tau}.
Recalling the
discussion of section 2, we see that the partition function $Z(S_{\M})$ can
be regarded as the regularized Euler number $\cs(\C)$ of $\C$. On the
other hand, from the previous section we already know that $Z(S_{\M})$
localizes onto the zeros of $s$, i.e. onto flat connections, and yields
the Euler number of $\M$ via the zero section of $T\M$ and the Gauss-Bonnet
theorem. Thus here we have yet another example in which the (ambiguous)
regularized
Euler number of an infinite dimensional vector bundle equals the
Euler number of a finite dimensional vector bundle (cf. (\ref{aj7})),
\be
\cs(\C)=\c(\M)\label{44}\;\;.
\ee
Moreover we know from \cite{tau} that for $M$ a homology three-sphere
(see section 4.2 for the definition) and $G=SU(2)$
\be
\cs(\C)=\lambda(M)\label{45}
\ee
where $\lambda(M)$ is the {\em Casson invariant} \cite{cas} of $M$
(in accordance with more recent work on the Casson invariant \cite{cap,wal}
we have dropped a factor of $1/2$ in the definition of $\lambda(M)$). This
identification as well as the implications of (\ref{44},\ref{45})
will be discussed further in sections 4.2 and 4.3.

In two dimensions the relevant part of the (reduced) action is
\bea
S_{\M}&=&\RC +\int_{M}BF_{A}+u(d_{A}\Sigma +[\p,\pb])
           +\Sigma*\Sigma +\pb d_{A}\c\nonumber\\
&=& \RC + \int_{\M}BF_{A}+u[\p,\pb]-\frac{1}{4}d_{A}u*d_{A}u+\pb d_{A}\c
    \label{46}\;\;.
\eea
As we have had to introduce a scalar superfield $\hat{B}$ in addition
to the superconnection $\hat{A}$, we expect the base space of the
bundle in question to be something like ${\cal A}\times\O{0}$ instead
of $\cal A$. The tangent space to ${\cal A}\times\O{0}$ at a point
$(A,u)$ is $T_{(A,u)}({\cal A}\times\O{0})=\O{1}\oplus\O{0}$, and
(\ref{46}) suggests the section $s(A,u)=(*d_{A}u,*F_{A})$. However,
this is strictly correct only if we add a $B^{2}$-term to the action
(\ref{46}) (which can be done in an $N=2$ invariant way). In the present
case (the `delta function gauge', cf. \cite{bbt,pr}) the correct geometrical
picture is obtained by integrating out $B$ and working directly with
the bundle $T{\cal A}|_{{\cal F}}\times T\O{0}$, where ${\cal F}=
\{A\in{\cal A}: F_{A}=0\}$ is the space of flat connections. The above
section now becomes $s(A,u)=(*d_{A}u,0)$. It gives rise to the
$s^{2}\sim\int_{M}d_{A}u*d_{A}u$ term of (\ref{46}) as well as to the
remaining two terms
\be
\int_{M}u[\p,\pb]+\pb d_{A}\c = -\int_{M}\pb \d (d_{A}u)\label{47}
\ee
which correspond to the third term $\sim\c ds$ of (\ref{eq:aj5}), with
$\d\widehat{=}\del_{\t}$ denoting the exterior derivative $\d A=\p,\d u=\c$
on ${\cal A}\times \O{0}$.

Away from reducible connections the zeros of this section are precisely
the flat connections: $s(A,u)=(0,0)\Leftrightarrow F_{A}=0,u=0$. It passes
down to a section of $T(\C)|_{\M}$ as $d_{A}^{*}(*d_{A}u)=0$ for
$A\in{\cal F}$. Note that $s$ does {\em not} define a section of $T\M$, but
rather of the normal bundle $N_{\M}$ of $T\M$ in $T(\C)|_{\M}$. Thus the
action (\ref{46}) of our two-dimensional $N=2$ topological gauge theory
can be regarded as the Mathai-Quillen representative of the
regularized Euler number $\cs(N_{\M})$ of the normal bundle $N_{\M}$.
Again, by the calculation of section 3.2, we know that this choice of
section regularizes this Euler number to be
\be
\cs(N_{\M})=\c(\M)\label{48}\;\;.
\ee
In the case of instantons in four dimensions everything runs as above
provided that $\O{0}$ is replaced by $\O{2}_{+}$.

\subsection{${\bf N=2}$ topological gauge theories, Floer cohomology, and
             supersymmetric quantum mechanics on ${\bf\C}$}

The considerations of the preceding section show that $N=2$ topological
gauge theories are based on the tangent bundle geometry of $\C$, in agreement
with the calculations of section 3.2 which exhibited a relation between
these theories and the Riemannian geometry of $\C$. This already makes
these theories much closer to supersymmetric quantum mechanics than, say,
Donaldson theory where the bundle in question \cite{aj} is not a natural
bundle (in the technical sense). However, the analogy is not yet perfect.
In order to gain a better understanding of the emergence of de Rham cohomology
in our $N=2$ models we will now construct supersymmetric quantum mechanics
on $\C$ along the lines of section 2 (i.e. {\it via} a section of the
tangent bundle of the loop space of $\C$). Alternatively, we can use the
Gauss-Codazzi form of supersymmetric quantum mechanics (section 2.3) to
get to the desired moduli space not via the zeros of a section but by means
of a supersymmetric delta function in the path integral.
For the space of gauge orbits
$\A{3}$ on a three-manifold $M$ we find that the resulting quantum mechanics
model for a particular choice of section
is precisely $N=1$ Donaldson theory on $M\times S^{1}$ which reduces to
the three-dimensional $N=2$ theory in the limit that the circle (time)
shrinks to zero (in particular, their partition functions are equal).
This gives a relation between the $N=2$ symmetry, de Rham theory on $\M_{3}$
and Floer cohomology. As all this is really just a reinterpretation
of the transition from the Hamiltonian \cite{at} to the Lagrangian
\cite{ewdon} description of Donaldson theory, we will not construct the
quantum mechanics action in detail (see \cite{btagqm}),
concentrating instead on the features relevant for us here. Analogous
considerations can be carried out, {\it mutatis mutandis}, in other
dimensions.

We begin with the space ${\cal A}^{3}$ of connections on a (trivial)
principal $G$-bundle over a three-manifold $M$ and would like to
interpret its loop space as the space of connections on some bundle
over $M\times S^{1}$. The first thing we should decide is whether
to work with $L{\cal A}^{3}$ (eventually modded out by $L{\cal G}^{3}$)
or with $L(\A{3})$. The difference between the two is that in the latter
case the connections are required to be periodic in time only up to a
gauge transformation, and it is that space we have to work with if we
are interested in non-trivial bundles on $M\times S^{1}$. To see this,
note that $L(\A{3})$ is not connected if there are large gauge transformations
on $M$,
\[\pi_{0}(L(\A{3}))=\pi_{0}({\cal G}^{3})\;\;,\]
and these can be used as clutching functions to construct non-trivial
bundles over the {\em mapping cylinder} $M\times S^{1}$ of $M$. $L(\A{3})$
represents the disjoint union of gauge equivalence classes of connections
on all isomorphism classes of bundles of $M\times S^{1}$ and
contains $L{\cal A}^{3}/L{\cal G}^{3}$ as its trivial component.
Since we are not going to worry about gauge fixing in the following
it is most convenient to work equivariantly on $L{\cal A}^{3}$.
However, the difference between the two spaces will occasionally be
crucial and we will draw attention to it when that occurs.

The obvious section to start off with is (as in section 2) $s_{0}(A)(t)
=\dot{A}(t)$. The corresponding Mathai-Quillen action (\ref{eq:aj5})
(i.e. the action (\ref{aj6}) with $x(t)$ replaced by $A(t)$)
has, however, still got a divergent partition function. In fact,
it would regularize the Euler number of $\A{4}=L{\cal A}^{3}/L{\cal G}^{3}$
to be $\cs(\A{4})=\c(\A{3})$, which is not yet well defined. An alternative
way of seeing this is to note that the dimensional reduction of this
action gives precisely what we called $S_{\A{3}}$ in section 3.2 - thus
our four-dimensional action still requires addition of a term corresponding to
$S_{\M}$.

The most natural way to try to do this is to change the section $s_{0}$ to
$s_{W}$ for some gauge invariant potential function $W$ on ${\cal A}^{3}$.
In the setting of section 2 this did not change the result: the partition
function of supersymmetric quantum mechanics is well defined and independent
of the choice of $W$, a statement equivalent to the classical formula
(\ref{morse}). In the present case the left hand side of (\ref{morse})
is not yet well defined, but we can make sense of it (regularize it further)
by {\em defining} it to be equal to the right hand side of (\ref{morse}) for
some choice of $W$ (cf. the discussion at the end of section 2.2).

A natural candidate for $W$ is the Chern-Simons functional
\be
W(A) = CS(A)\equiv\int_{M}AdA+\frac{1}{3}A[A,A] \label{49}\;\;.
\ee
$CS(A)$ is not quite gauge invariant (it changes by a constant proportional
to the winding number under large gauge transformations) but its
derivative is, and this is sufficient for our purposes. This choice of
potential defines the section $s_{CS}(A)(t)=\dot{A}(t)+*F_{A(t)}$ of
$T(L{\cal A}^{3})$.

By general results on
supersymmetric quantum mechanics \cite{pr} (or explicit calculation)
one finds that the partition function of the action corresponding to
$s_{CS}$ localizes onto the zeros of $s_{CS}$ (in the present case
possibly modified by terms required for four-dimensional gauge invariance
which is not guaranteed by the Mathai-Quillen formalism - this will not
affect any of our conclusions).
If $A(t)$ is periodic in $t$ (this means that we are in the topologically
trivial sector) the same `squaring argument' as in section 2
shows that these are precisely the time-independent flat connections on
$M$. In the topologicaly non-trivial sectors the `squaring argument'
fails and there are non-trivial solutions to the equation (the
gradient equation of the Chern-Simons functional)
\be
\frac{d}{dt}A(t)=-*F_{A(t)}\label{50}\;\;.
\ee

Equation (\ref{50}) (usually read as an equation on $M\times{\bf R}$)
is nothing other than the instanton equation in the $A_{0}=0$ gauge and
plays a fundamental role in defining the relative Morse
indices of Floer's instanton
(co)homology \cite{floer,at} and also provides the
link between the three-dimensional Floer cohomology groups and the
four-dimensional Donaldson invariants \cite{don} associated with
the moduli spaces of instantons (see also \cite{at,pr} for the
definition of these invariants and their relation with Floer theory).

This already suggests that we have just reinvented the wheel and that the
four-dimensional topological gauge theory we have constructed here is
nothing other than Donaldson theory. That this is indeed the case
becomes immediately obvious by noticing that
the Hamiltonian of the corresponding
quantum mechanics action is the Laplacian of $\d_{CS}$,
the exterior derivative
$\d$ on ${\cal A}^{3}$ twisted by the Chern-Simons functional, and hence
precisely the Hamiltonian of Donaldson theory \cite{at,ewdon}.
Alternatively this can of course be seen directly at the level of the action
\cite{btagqm}. There are
some things that we ought nevertheless to check to be sure that all the
pieces fit. Firstly, we know from the four-dimensional standpoint
that the partition function of the Donaldson
theory vanishes if the index of the deformation complex (the formal
dimension of the instanton moduli space) is not
zero. An essential ingredient in the construction is then that this
index be zero. On the other hand, to get sensible results
for the three dimensional theory the
only sector in the four dimensional theory theory which contributes must
be the one with trivial second Chern class. We will now see how these two
requirements take care of each other. The index is equal to \cite{ahs}
\be
p_{1} \, - \, \frac{1}{2} dimG ( \, \chi(M_{4}) \, + \, \sigma(M_{4}) )\;\;.
\label{dim}
\ee
Here $p_{1}$ is the first Pontryagin number of the adjoint bundle $ad\,P$
(equal to $8k$, $k$ the instanton number, for $G=SU(2)$) and
$\sigma(M_{4})$ is the signature of $M_{4}$. This is the signature of the
intersection form on $H^{2}(M_{4},{\bf Z})$ or the number of self-dual minus
the number of anti-self-dual harmonic two-forms on $M_{4}$. The
Euler characteristic of a four-manifold of the form $M_{4}=M_{3}
\times S^{1}$ is zero as $\chi(M_{3} \times S^{1}) = \chi(M_{3}) \chi(S^{1})
= 0$. By the same multiplicative property of the signature
\cite[Theorem 8.2.1]{hirz} $\sigma(M_{3}\times S^{1})$
also vanishes. Explicitly this can be seen as follows:
the K\"unneth formula tells us that
$H^{2}(M_{3} \times S^{1};R)$ is isomorphic to $H^{1}(M_{3};R) \oplus
H^{1}(M_{3};R)$, and if $\{ \omega_{i} \}$ form a basis for $H^{1}(M_{3};R)$
then (symbolically)
a basis for $H^{2}(M_{3} \times S^{1};R)$ is $\{ \omega_{i} \oplus
\omega_{i} , \omega_{i} \oplus - \omega_{i} \}$. The first entry forms a
basis for the space of
self-dual harmonic two forms $H^{2}_{+}$ while the second is a
basis for that of anti self-dual harmonic two forms $H^{2}_{-}$: they
necessarily have the same dimension\footnote{Alternatively we note that
$Tr \int RR$ vanishes with the product metric which implies the vanishing
of $\sigma(M_{3}\times S^{1})$ by the Hirzebruch signature theorem.}.
hence we find that the index
is non-zero for all $p_{1}\neq 0$. Notice
that while the dimension formula (\ref{dim}) tells us that the instanton
moduli space is formally zero for $p_{1}=0$
it tells us nothing about the dimension of the space of flat connections.
In fact, the index of the flat connection deformation complex is the sum
of the instanton and anti-instanton indices for $p_{1}=0$ and is given
by $-dimG \c(M_{4})$ (the index of the twisted de Rham complex) which is
zero for $M_{4}=M_{3}\times S^{1}$.

This also settles the
question raised above whether we should work with $L(\A{3})$ or
$L{\cal A}^{3}/L{\cal G}^{3}$: the theories are identical in the
sector with $p_{1}=0$, which is (according to the above) the only one that
will contribute to the partition function, so for our purposes
both alternatives are equivalent.
In the topologically trivial sector
the partition function reduces to an integral over the moduli space
$\M_{3}(M,G)$ for an arbitrary three-manifold $M$.
There the twisted exterior derivative
$\d_{CS}=\d +\int_{M}F_{A}\d A$ reduces to the
ordinary exterior derivative, the Hamiltonian to the
Laplacian on $\M_{3}$, and the partition function is (independently
of the radius of the circle)
the Euler characteristic $\c(\M)$ of the de Rham complex
of $\M$.

That this agrees with the partition function of the three-dimensional
$N=2$ theory is no coincidence. In fact, we can expand all fields
in Fourier modes along the circle. Integrating out the non-constant modes
the resulting three-dimensional action is precisely the action
constructed in section 3.2. In the light of our previous considerations
and those of \cite{ewtop,brt} (the three-dimensional theory is the
dimensional reduction of Donaldson theory) this is not very surprising
and the calculational details can be found in \cite{btagqm}.
This finally also establishes the sought-for direct relation between this
$N=2$ topological gauge theory and supersymmetric quantum mechanics on
spaces of connections.

Another part of the puzzle that fits in
place is that ignoring time derivatives the $N=1$ symmetry of Donaldson
theory extends to an $N=2$ symmetry which is enjoyed by the Hamiltonian
(ie. the Lagrangian of the three dimensional theory) just as we found
when embedding $S^{2}$ into $R^{3}$ in section 2.3.
Indeed, as $H \sim \int T_{00}$ and $T_{00} = \{ Q, V_{00} \}$ for some
$V_{00}$ (by the fundamental property $T_{\alpha\beta}=\{Q,V_{\alpha\beta}\}$
of topological field theories \cite{ewdon})
we could imagine that $H = \{ Q, \bar{Q} \}$, where $\bar{Q}$
is nilpotent and leaves the Hamiltonian invariant. That this is indeed
the case was established by Witten in \cite{ewdon}.

Returning to the three dimensional discussion, let us momentarily assume that
$M$ is a homology three-sphere, i.e. an orientable closed
three-manifold with $H_{1}(M,{\bf Z})=0$, and that the gauge group is
$G=SU(2)$.
In that case flat connections are isolated and (apart from
the trivial connection) irreducible (in fact, a reducible connection,
defined by a reducible element of $Hom(\pi_{1}(M),G)$, would factor through
to an element of $Hom(\pi_{1}(M),U(1))\approx H_{1}(M,U(1))=0$).
In this setting the Floer cohomology groups, the cohomology groups
of the twisted exterior derivative
$\d_{CS}$ are
well defined and coincide (as in ordinary supersymmetric quantum mechanics)
with the ground states of the above Hamiltonian. In particular, therefore,
the partition function of this theory is (ignoring problems with the
trivial connection)
the Euler characteristic $\c_{F}(M)$
of the Floer complex which is known \cite{at,tau} to be
\be
\c_{F}(M)=\lambda(M)\;\;. \label{51}
\ee
Note that the calculation of the Euler characteristic of the Floer complex
requires only the topologically trivial sector (flat connections) although
the definition of the individual instanton homology groups depends crucially
on all the topologically non-trivial sectors (instantons). This is entirely
analogous to ordinary supersymmetric quantum mechanics on a manifold $M$: the
Euler number $\c(M)$ can be calculated in terms of the fixed points of some
vector field alone whereas instanton paths connecting these fixed points enter
into the computation of the homology groups of $M$ \cite{ewqm}.

The consequences of the intriguing equations $\c_{s}(\A{3})=\c(\M_{3})$
(\ref{44}), $\c_{s}(\A{3})=\lambda(M)$ (\ref{45}) and $\c_{F}(M)=\lambda(M)$
(\ref{51}) will be explored in the following section, after we have recalled
the definition and some properties of the Casson invariant.

\subsection{The Casson invariant and its generalization}

In this section we will deal exclusively with the three-dimensional theory
defined by the action (\ref{28})+(\ref{38}). We have seen above that
formally the partition function of this theory yields the Euler characteristic
of the moduli space $\M_{3}$ of flat connections via the Gauss-Codazzi
equations
and the Gauss-Bonnet theorem. On the other hand we will see below that (again
formally) the partition function is the Casson invariant if $M$ is an
integral homology three-sphere. Of course, these two observations taken
together immediately suggest a generalization of the Casson invariant to
arbitrary three-manifolds. But in order to substantiate this suggestion, there
are some problems that need to be overcome at a purely mathematical level
before one can try to assert whether $\c(\M)$ is a meaningful and useful
generalization of the Casson invariant. In particular, one needs to

\noindent a) define what one means by $\c(\M)$ when $\M$ is not a smooth
manifold but perhaps (at best) an orbifold stratification (in the sense
of Kirwan \cite{kirwan}), and

\noindent b) compare candidate definitions of $\c(\M)$ with already
existing extensions of the Casson invariant to certain more general
classes of three-manifolds (rational homology spheres \cite{cap,wal},
homology lens spaces \cite{bl}).

We have no definite solutions to offer to these problems but we will provide
some background information and preliminary suggestions below
which we believe will play a role in the resolution of these issues.

In addition to these mathematical issues (which are completely independent
of the field theoretic considerations by which we were led to them) there
are problems with the field theoretic realization of these topological
(differential) invariants. In particular, in order to be able to assert that
the partition function really calculates the Casson invariant (in the
simplest case of homology spheres) or the Euler number, one needs to

\noindent c) come to terms with the contributions from the trivial connection
and other reducible connections (we will be more precise about what is really
required in the case of integral or rational homology three-spheres below).

Of course, c) is not completely independent of the issues a) and b) above
in the sense that knowing the correct way of treating reducible connections
and the singularities they give rise to on the mathematical side may
provide a prescription or guideline for dealing with them in the path integral.
Again we have no immediate solutions to offer, but we will make some
comments on this part of the story as we go along.
At first, however, our considerations will be formal.

Let us again assume for the time being that
$M$ is a homology three-sphere.
There are then no $\p$ and $\pb$ zero modes (i.e. no non-trivial solutions
$\p_{A}$ to the equations $d_{A}\p = d_{A}*\p =0$, etc.) and the
partition function will simply reduce to a sum of contributions from the
points of $\M$ (cf. (\ref{36})), which - by supersymmetry - are plus or
minus one,
\be
 Z(S_{\M}) = \sum_{\M}\pm 1\label{52}
\ee
(the contribution of the trivial connection is ill-defined at this point
and is assumed to be excluded from the sum (\ref{52}) until further notice).
A look at the action $S_{\M}^{0}=\int_{M}BF_{A}+\pb d_{A}\p$
reveals that the {\em relative} signs are determined by
the (mod 2) spectral flow of the operator $d_{A}$, the same spectral flow
that defines the relative Morse indices of Floer homology \cite{floer,at}.
Therefore $Z(S_{\M})$ equals the
Euler characteristic $\c_{F}(M)$ of the Floer complex.
Since $d_{A}$ is the Hessian of the Chern-Simons functional whose first
derivative defines the gradient vector field $*F_{A}$ on ${\cal A/G}$,
we see rather directly that $Z(S_{\M})$ can be regarded as
defining the Euler number $\cs(\C)$, as we of course already know more
generally from section 4.1.

It is a result of Taubes \cite{tau}
that this topological invariant agrees (possibly
up to a sign) with
the Casson invariant \cite{cas} $\lambda(M)$,
\be
Z(S_{\M}) = \lambda(M) \label{53}
\ee
(again, provided that the trivial connection is excluded from the sum).
Actually Taubes also fixes the absolute sign. This requires considerations
involving perturbations of the trivial connection, and we will come back to
this below.
$\lambda(M)$ is a very powerful (differential) integer-valued invariant of
homology three-spheres which
generalizes the classical Rohlin invariant (with which it agrees mod 2) and has
already led to many interesting results in low-dimensional topology \cite{cas}.
One of the most striking results is perhaps a very simple proof of the
existence of non-triangulable four-manifolds (e.g. Freedman's $E_{8}$
\cite{freedman}).

Casson's original definition of $\lambda(M)$ was somewhat different, involving
Heegard splittings of $M$ along a Riemann surface $\Sigma_{g}$, and
intersection
theory in $\M(\Sigma_{g},SU(2))$. We will now show how his definition can be
recovered from the path integral point of view (this is taken from \cite{pr}).
Imagine splitting $M$ along
a Riemann surface $\Sigma_{g}$, i.e. $M=M_{1}\sharp_{\Sigma_{g}}M_{2}$, where
$M_{1}$ and $M_{2}$ are handlebodies (solid Riemann surfaces). Then
- according to the general principles of quantum field theory - the path
integral over connections on the manifold $M_{1}$ with boundary
$\partial M_{1}=\Sigma_{g}$ will define a wave function
$\Psi_{1}$  having support on those flat connections on $\Sigma_{g}$ which
extend to flat connections on $M_{1}$, i.e. on the Lagrangian submanifold
$\M(M_{1},SU(2))$ of the symplectic manifold
$\M(\Sigma_{g},SU(2))$. Likewise the path integral over connections on $M_{2}$
will produce a wave function $\Psi_{2}$
having support on $\M(M_{2},SU(2))\subset
\M(\Sigma_{g},SU(2))$. The partition function $Z(M)$ can then be computed
as the scalar product
\be
Z(M)=\int_{\M(\Sigma_{g},SU(2))} \Psi_{1}^{*}\Psi_{2} \label{54}\;\;,
\ee
and evidently only receives contributions from flat connections on $\Sigma_{g}$
which extend
to both $M_{1}$ and $M_{2}$ or - in other words - from flat connections on $M$.
By our assumption that $M$ is a homology three-sphere this implies, that
(\ref{54}) is a sum over the points of $\M(M,SU(2))$.
The key point in Taubes' work is to show that the relative algebraic
intersection numbers of $\M(M_{1},SU(2))$ and $\M(M_{2},SU(2))$
in $\M(\Sigma_{g},SU(2))$ can be
determined  from the spectral flow of $d_{A}$.  Then one gets (denoting
the total intersection number in $\M$ by $\sharp_{\M}$)
\be
\lambda(M)=
\sharp_{\M(\Sigma_{g},SU(2))}(\M(M_{1},SU(2)),\M(M_{2},SU(2)))
\label{55}\;\;,
\ee
which is precisely Casson's original definition (up to the factor of $1/2$
mentioned above).

For $M$ an integral homology three-sphere this is well defined as
\[\M(M_{1},SU(2))\cap\M(M_{2},SU(2))\subset\M(\Sigma_{g},SU(2))\]
does not contain any singular points
of $\M(\Sigma_{g},SU(2))$ apart from the trivial connection $\Gamma$. This is
due to the fact that in two dimensions singularities of $\M$ arise only from
reducible connections \cite{gol} and that the only reducible connection
contributing to the intersection is $\Gamma$ if $M$ is a homology three-sphere.
In higher dimensions singularities may have other origins as well and we will
briefly have to come back to this below.

The overall sign of (\ref{55}) is determined by a careful comparison of the
orientations of the manifolds $M_{k}$ and $\Sigma_{g}$ and their moduli
spaces. In Taubes' gauge theoretic approach, on the other hand, the trivial
connection $\Gamma$ determines the sign of $\lambda(M)$ in the following way.
Recall that the individual relative conributions $\pm 1$ to the sum
(\ref{52}) are defined by the (mod 2) spectral flow of the operator $d_{A}$
between the irreducible flat connections. In the absence of any preferred
connection this still leaves undetermined the overall sign. Now $\Gamma$ is
such a preferred connection. Unfortunately, the spectral flow between $\Gamma$
and any irreducible connection is ill-defined as $d_{\Gamma}$ has a triple
zero eigenvalue. The way to get around this difficulty is to perturb $\Gamma$
slightly,
\[\Gamma\ra\Gamma +\epsilon a\;\;,\;\;\;\;\;\;\epsilon >0\;\;,\]
so that the perturbed operator (or, rather, an extension thereof,
cf. \cite{tau}) has three small $O(\epsilon^{2})$ non-zero eigenvalues.
This operator and the spectral flow between $\Gamma +\epsilon a$
and some irreducible flat reference connection $A_{0}$
can then be used to assign a
well-defined (perturbation independent) sign to $\lambda(M)$.

{}From the path integral point of view it is not obvious how the trivial
connection $\Gamma$ should be dealt with. The lesson to be learned from
this discussion is that the correct procedure is neither to try to assign
a sign to it contributing $\pm 1$ to the sum (\ref{52}) (in fact,
$\lambda(M)=0$ if $\pi_{1}(M)=1$) nor to simply drop it, but to deal with
it in such a way that it only effects the overall sign of the partition
function. It appears likely that precisely this will happen when one
BRST gauge fixes the additional finite-dimensional $SU(2)$ symmetry appearing
at $\Gamma$ in the action (\ref{28})+(\ref{38}) gauge fixed at $A_{0}$.
Anyway, this and the treatment of the trivial connection in all other
(topological) gauge theories is certainly an important issue that still
remains to be understood.

In recent years some effort has gone into generalizing the Casson invariant
to other groups $G$ or to more general classes of three-manifolds (see e.g.
\cite{cap,wal,bl}). One is then inevitably confronted with the presence of
non-trivial reducible flat connections. In Casson's approach this is
problematic because $\M(M_{1},G)\cap\M(M_{2},G)$ now meets the singularities
of $\M(\Sigma_{g},G)$ and the definition of the intersection numbers requires
more care. Alternatively, in Taubes' approach (which has not yet been worked
out in a more general setting) a more delicate perturbation theory would be
required to deal with the zero modes of $d_{A}$ at these points.

Important progress was made by Walker \cite{wal} who extended the definition
of the Casson invariant to rational homology three-spheres
($H_{*}(M,{\bf Q})=H_{*}(S^{3},{\bf Q})$), and by Cappell, Lee and Miller
\cite{cap} who generalized it to arbitrary semisimple Lie groups $G$.
Moreover, in the `Note Added' to \cite{cap}, they announce an extension
of $\lambda(M)$ to arbitrary orientable closed three-manifolds. In all these
generalizations, $\lambda(M)$ is no longer necessarily an integer.

In the case considered by Walker (which, for simplicity,
is the only one we will consider here)
the contribution from a $U(1)$-reducible flat connection
can be expressed either in terms of (secondary) characteristic classes
of the normal bundle of the space of reducible flat connections in
$\M(\Sigma_{g},SU(2))$ \cite{wal}, or in terms of Maslov indices,
or in terms of the line integral of the Bismut-Freed connection \cite{bf}
on a certain determinant line bundle over the space of reducible connections
\cite{cap}. In each case the precise definition involves paths in
$\M(M_{1},G)$ and $\M(M_{2},G)$ from
$\Gamma$ to the non-trivial reducible connections. All these data are rather
natural from a path integral point of view, and the proper treatment of
reducible connections in the corresponding topological gauge theory should
be guided by the requirement that either of these expressions is reproduced.
Again progress along these lines would constitute a major step forward in
our understanding of reducible connections in (topological) gauge theories.

{}From a different angle we have seen at various points in this paper that
it is natural to propose the Euler characteristic $\c(\M(M,G))$ as a
generalization of the Casson invariant. For integral homology spheres
these two definitions coincide (as the spectral flow of $d_{A}$ indeed
measures the relative tangent space (point) orientations). Moreover,
the same topological theory that formally gives us the Casson
invariant if the underlying three-manifold $M$ is a homology sphere
formally computes the Euler number of $\M(M)$ via the Gauss-Codazzi
equations and the Gauss-Bonnet theorem when the dimension of $\M(M)$ is
non-zero. This permits us to identify the (ambiguous) regularized Euler
number of $\A{3}$ (in the sense of Atiyah-Jeffrey and Taubes)
with the Euler number of $\M(M)$.

This suggestion by itself does, of course, not solve any of the technical
problems inherent in the definition of $\c(\M)$ for the types of spaces
arising as moduli spaces of flat connections on three-manifolds. Nevertheless,
we hope that this proposal is concrete enough to be useful and perhaps
guide future investigations, as the problem is now more specifically that
of finding a `good' definition of $\c(\M)$. The difficulties one encounters
when trying to find such a definition are all related in one way or
another to the fact that the singularity-structure of the spaces $\M(M,G)$,
which we will in the following think of as
\be
\M(M,G)=Hom(\pi_{1}(M),G)/G\;\;,\label{56}
\ee
(the quotient action is by conjugation) is not well understood.

For instance, in three dimensions singularities are not only due to
reducible representations of $\pi_{1}(M)$ but also to the relations
satisfied by the generators of $\pi_{1}(M)$. Moreover, again in
contrast with two dimensions, there is no reason to expect the spaces
$\M(M,G)$ to be symplectic in general. This means, in particular,
that the study of the singularities of $\M(M,G)$ does not reduce to that
of singularities of moment maps, for which powerful techniques
and general results would have been available.

A related (but more fundamental) difficulty is that the set of all fundamental
groups of three-manifolds is not well understood. In fact, this set
is not even algorithmically recognizable in the set of all finitely generated
groups \cite{hempel,fomenko} so that one cannot expect any general
theorems describing the structure of the spaces (\ref{56}). It is known that
the infinite fundamental groups have at most 2-torsion (and are torsion
free for orientable manifolds) and that the possible Abelian groups are
exhausted by the list
\[{\bf Z}\;\;,\;\;{\bf Z}^{2}\;\;,\;\;{\bf Z}^{3}\;\;,\;\;{\bf Z}_{p}\;\;,
\;\;{\bf Z}\oplus{\bf Z}_{2}\;\;,\]
and at least in the latter case it is possible to work out the structure of
$\M(M,G)$ explicitly. But this nevertheless still leaves the whole menagerie
of finite non-Abelian groups unaccounted for.

More generally, the spaces (\ref{56}) have been studied in the context of
representation varieties of finitely generated groups (see the monograph
\cite{lm} and the contributions in \cite{gm}). From these investigations
it is known that for certain types of finitely generated
groups $\pi$ (e.g. nilpotent) the irreducible subvariety of $Hom(\pi,G)/G$
is non-singular. However, again little is known about the nature of the
singularities of $Hom(\pi,G)/G$ away from irreducible representations.

Thinking now more concretely about defining the Euler number of singular
spaces we want to mention the encouraging result that there is a Gauss-Bonnet
theorem for V-manifolds (orbifolds) \cite{satake} which calculates the
virtual Euler characteristic of an orbifold as defined e.g. in
\cite{satake}-\cite{brown}
in various contexts. The virtual Euler number is different from the
topological Euler number of an orbifold and is no longer necessarily an
integer if the orbifold is not a smooth manifold. This checks with the
properties of the Casson invariant away from integral homology spheres.

If the moduli space $\M\subset\C$ is an orbifold then
the metric induced on $\M$ by the metric on $\C$ will be an orbifold metric
(in the sense of \cite{satake}) and therefore the evaluation of the partition
function (i.e. of the Gauss-Bonnet integrand) will give rise to the virtual
(orbifold) Euler number of $\M$. On the other hand it is not so clear what
one is calculating in the case of stratified spaces.  We are not aware of
a similar generally accepted definition of the Euler number of a stratified
manifold apart from the purely topological version in terms of equivariant
Betti numbers \cite{ab} which is unlikely to capture the relevant
information in the present case (cf. the generalizations \cite{cap,wal}
of the Casson invariant mentioned above).

Additional circumstantial evidence in favour of our suggestion could be
provided by showing that - at least formally - the Euler number
$\c(\M(M,G))$ has in general properties similar to those
satisfied by the Casson invariant
$\lambda(M,G)$ in the case of integral or rational homology spheres,
like additivity under connected sums,
\be
\lambda(M_{1}\sharp M_{2},G)= \lambda(M_{1},G)+\lambda(M_{2},G)\;\;,\label{57}
\ee
or its behaviour
\be
\lambda(-M,G)=(-)^{dim(G)}\lambda(M,G)\label{58}
\ee
under reversal of orientation of $M$. These properties may however have
to be relaxed anyway when moving away from (rational) homology spheres.
For instance, property (\ref{57}) is satisfied by the extension of the
Casson invariant to homology lens spaces considered by Boyer and Lines
\cite{bl} if the orders of the first homology
groups ${\bf Z}_{p_{k}}$ of $M_{k}$ are co-prime, but not in general.

One of the most important properties of the Casson invariant (apart from
being a differential invariant) is its nice behaviour under Dehn
surgery on knots and its relation with the Alexander polynomial. This is
a property that one may not wish to give up, but unfortunately also one
that seems to be rather difficult to prove for $\c(\M)$.

Whatever the outcome of these
investigations will be, we hope that thinking
in terms of traditional differential-geometric concepts will contribute
to the understanding of the Casson invariant and its generalizations.

\section{Concluding Remarks: the Penner Model and Other Open Questions
and Generalizations}

In this paper we have drawn together a number of threads to construct a
topological gauge theory with the property that its partition function
is the Euler number $\c(\M)$ of some given finite dimensional moduli
space $\M$ of connections. Among these threads were supersymmetric
quantum mechanics, its relation with the Mathai-Quillen formalism,
the Gauss-Codazzi equations for $\M\subset\C$, a superfield construction
of $N=2$ topological gauge theories, and the Atiyah-Jeffrey interpretation
of topological field theories.

Along the way we have also obtained some results which are potentially
interesting outside the context of topological field theories as well.
In particular, we have introduced a new kind of supersymmetric quantum
mechanics based on (or: deriving) the Gauss-Codazzi equations of classical
Riemannian geometry. We believe that we have also clarified the concept
of the regularized (Mathai-Quillen) Euler number $\cs(E)$ of an infinite
dimensional vector bundle $E$, introduced by Taubes and Atiyah-Jeffrey,
by showing that under very general conditions $\cs(E)$ can be identified
with the rigorously and unambiguously defined Euler number of some
finite dimensional vector bundle. Combined with the fact that the Casson
invariant of a homology three-sphere can be interpreted as $\cs(\C)$ and
with the observation that the partition function of one and the same
topological gauge theory (formally) yields either the Casson invariant
or the Euler number $\c(\M)$ of the moduli space of flat connections,
depending on the dimension of $\M$, this led us to suggest $\c(\M)$ as a
generalization of the Casson invariant to other classes of three-manifolds.

In the previous section we have mentioned some of the technical problems
one encounters when attempting to a) make this suggestion more precise
from a purely mathematical point of view, and b) put the field theoretic
considerations on a slightly more rigorous footing. We pointed out that
the work done on the Casson invariant and its generalizations may provide
a prescription, or at least a guideline, for dealing with the trivial and
other reducible connections in the path integral.

In addition to these technical questions there are a number of other open
problems and avenues for future research. In particular, we want to draw
attention to the possibility of constructing a topological counterpart of
the Penner matrix model \cite{pen}. This is a hermitian
matrix model whose
partition function calculates the virtual (orbifold) Euler characteristic
of the moduli space of Riemann surfaces of genus $g$ \cite{hz}. Inspired
by the relation between $2d$ gravity and the continuum limit of matrix
models, which has led to a dramatic improvement in the understanding
of non-critical string theories in recent years,
Distler and Vafa \cite{dv} investigated the question whether there is a
$2d$ gravity theory describing the continuum limit of the Penner matrix model.
By considering the scaling behaviour they concluded that such a theory
would have to have central charge $c=1$, in agreement with the explicit form
of the free energy of the continuum Penner model which is related to that
of a string propagating on a circle at the self-dual radius.

In analogy with Distler's observation \cite{distler} that topological
gravity \cite{lpw} is a bosonized form of Liouville theory coupled to
$c=-2$ matter one may speculate on the existence of a topological
field theory describing Liouville theory coupled to the $c=1$ model
conjectured by Distler and Vafa. Such a topological theory would be
characterized by the property that its partition function is the Euler
number of moduli space, precisely the property shared by the topological
models discussed in this paper.

Recalling that Teichm\"uller space can
be regarded as a particular component of the moduli space of flat
$PSL(2,{\bf R})$ connections in two dimensions \cite{gol}
the point of departure could thus be the two-dimensional topological
gauge theory of section 3.2.
Alternatively one can write down directly an $N=2$ version of the
Labastida-Pernici-Witten model \cite{lpw} in which the fundamental
field is the metric and not a $PSL(2,{\bf R})$ connection. In either
approach it remains, however, to implement the mapping class group
which acts on Teichm\"uller space and whose orbit space is the moduli
space of surfaces. Having accomplished this one could then, turning around
the calculation of Distler, determine the $c=1$ matter system by
demanding that its coupling to Liouville theory is equivalent to
the thus constructed topological theory. It would be interesting
to understand the relation with the gauged $SU(2)/U(1)$
WZW model at level $k=-3$ which was recently shown by Witten \cite{ewwzw}
to also calculate the Euler number of moduli space.

Another open problem in this context is
whether this or other $N=2$ models in two dimensions can be described as
`twisted' $N=4$ superconformal models. It is known that
$N=2$ superconformal models can be twisted to $N=1$ topological theories
\cite{ewtg,ey} and that twisted $N=3$ theories describe
supersymmetric $N=1$ topological theories while twisted superconformal
$N=4$ theories appear to describe non-supersymmetric topological $N=2$
theories \cite{yoshii}. We hope to report on progress
along these lines in the future \cite{bjnt}.

More immediate generalizations of the models discussed in this paper are
e.g. supersymmetric extensions or a reformulation of the three-dimensional
model of flat connections on $R^{3}$ and the hyperbolic three-plane to
describe moduli spaces of monopoles (see \cite{brt,bg}).
It is also possible to construct
topological $N=2$ sigma models and these have the expected property of
describing the Riemannian geometry of spaces of sigma model instantons.
The details are in either case not too difficult and are left to the reader.

Finally we want to mention that it is possible to add a Chern-Simons term
to the three-dimensional $N=2$ action of flat connections.
The resulting action has still
got a topological $N=2$ symmetry (albeit slightly different from the one
considered in this paper). The partition function is now not the
number of flat connections counted with signs but rather a sum over flat
connections weighted by signs and phases (the exponential of the Chern-Simons
invariant of the flat connection).  This begs the question, with which we
conclude this paper, if this is an
interesting refinement of the Casson invariant of homology spheres.
\vspace*{.4in}
\begin{center}
{\bf Acknowledgements}
\end{center}
\noindent
We are grateful to C.~H.~Taubes for encouraging correspondence and
M.~S.~Narasimhan for his interest. We acknowledge the financial
support of the Stichting voor Fundamenteel Onderzoek der Materie
(FOM), and the Bundesministerium f\"ur Forschung und Technologie
under contract 06MZ760.

\end{document}